# Recent advances in iron-based superconductors toward applications

Hideo Hosono[a, b, *], Akiyasu Yamamoto[c], Hidenori Hiramatsu[a, b], and Yanwei Ma[d]

[a] Laboratory for Materials and Structures, Institute of Innovative Research, Tokyo Institute of Technology, Mailbox R3-3, 4259 Nagatsuta-cho, Midori-ku, Yokohama 226-8503, Japan
[b] Materials Research Center for Element Strategy, Tokyo Institute of Technology, Mailbox SE-1, 4259 Nagatsuta-cho, Midori-ku, Yokohama 226-8503, Japan
[c] Department of Applied Physics, Tokyo University of Agriculture and Technology, 2-24-16 Naka-cho, Koganei-shi, Tokyo 184-8588, Japan
[d] Key Laboratory of Applied Superconductivity, Institute of Electrical Engineering, Chinese Academy of Sciences, Beijing 100190, People's Republic of China

*Corresponding author: H. Hosono (hosono@msl.titech.ac.jp)

Abstract

Iron with a large magnetic moment was widely believed to be harmful to the emergence of superconductivity because of the competition between the static ordering of electron spins and the dynamic formation of electron pairs (Cooper pairs). Thus, the discovery of a high critical temperature ($T_c$) iron-based superconductor (IBSC) in 2008 was accepted with surprise in the condensed matter community and rekindled extensive study globally. IBSCs have since grown to become a new class of high-$T_c$ superconductors next to the high-$T_c$ cuprates discovered in 1986. The rapid research progress in the science and technology of IBSCs over the past decade has resulted in the accumulation of a vast amount of knowledge on IBSC materials, mechanisms, properties, and applications with the publication of more than several tens of thousands of papers. This article reviews recent progress in the technical applications (bulk magnets, thin films, and wires) of IBSCs in addition to their fundamental material characteristics. Highlights of their applications include high-field bulk magnets workable at 15–25 K, thin films with high critical current density ($J_c$) > 1 MA/cm$^2$ at ~10 T and 4 K, and an average $J_c$ of $1.3 \times 10^4$ A/cm$^2$ at 10 T and 4 K achieved for a 100-m-class-length wire. These achievements are based on the intrinsically advantageous properties of IBSCs such as the higher crystallographic symmetry of the superconducting phase, higher critical magnetic field, and larger critical grain boundary angle to maintain high $J_c$. These properties also make IBSCs promising for applications using high magnetic fields.

1. Introduction

Superconductivity, a clear-cut quantum phenomenon of zero electrical resistivity and perfect diamagnetism, was discovered in 1911 for mercury [1]. An extensive search for novel superconductors, in particular materials with higher critical temperature ($T_c$), has been performed since then. Figure 1 plots the $T_c$ values of the different classes of superconducting materials at ambient pressure as a function of the year that the superconductor was discovered or the max $T_c$ was reported. Superconducting materials can be classified into three categories based on the parent materials: metal-based system, copper oxides (cuprates), and iron pnictides/chalcogenides (which are called iron-based superconductors (IBSCs)). Each superconductor class is described in chronological order [2].

Representative metal-based superconductors have a cubic crystal structure called the A15 structure, which is shown in the inset of Fig. 1. $T_c$ was slowly increased by tuning the constituting elements in this structure from $V_3Si$ in 1953 to $Nb_3Ge$ in 1973. This advance is largely attributed to Berndt Matthias who discovered ~1,000 new superconductors. Matthias formulated his famous six rules for a successful search for new superconductors [3]: (1) high symmetry is good; cubic symmetry is the best, (2) a high density of electronic states is good, (3) stay away from oxygen, (4) stay away from magnetism, (5) stay away from insulators, and (6) stay away from theorists. These empirical 'rules' appeared to work well in the exploration of metal-based superconductors, as evidenced by the $T_c$ enhancement in A15-type metal alloys. The dominant mechanism in this material system is a phonon-mediated mechanism, in which dynamic pairing of conducting electrons called "Cooper pairs" (a requisite for the emergence of superconductivity) is caused by coupling with lattice vibrations. Superconductors dominated by this phonon mechanism are called conventional superconductors because this mechanism corresponds to the prototype of BCS theory proposed in 1957 by Bardeen, Cooper, and Schrieffer who first succeeded in elucidating a microscopic mechanism for superconductivity. The maximum $T_c$ for this type of superconductors saturated near the 23 K value reported for $Nb_3Ge$ in 1973. Many researchers believed that this maximum $T_c$ was an intrinsic limit controlled by the mechanism based on electron–phonon coupling (BCS theory), calling it 'the BCS wall'.

In 1986, a new superconductor with $T_c$ exceeding 'the BCS wall' was reported by Johannes Georg Bednorz and Alex Müller who discovered a $T_c^{onset}$ of 31 K in the Ba–La–Cu–O system with a layered perovskite structure [4]. The title of this groundbreaking paper published was 'Possible High $T_c$ Superconductivity in the Ba–La–Cu–O system' and the solid confirmation of this bulk superconductivity based on measurements of the Meissner effect was obtained by a research group at the University of Tokyo led by Shoji Tanaka and Koichi Kitazawa at the end of that year [5]. These findings kindled unprecedented worldwide extensive research into superconductors. This enthusiasm originated from the fact that a $T_c$ value higher than the BCS wall was discovered in a material that completely disobeyed the Matthias rules; i.e., the material was an oxide with non-cubic symmetry and the parent material before carrier doping was a Mott insulator with anti-ferromagnetism (AFM) arising from $Cu^{2+}(3d^9)$ [6].

In March 1986, Maw-Kuen Wu et al. reported superconductivity at 93 K in the Y–Ba–Cu–O compound system [7]. This material was the first with $T_c$ higher than 77 K (boiling point of liquid $N_2$), and practical application of superconductivity has been greatly expected since then. Subsequently, higher $T_c$ materials were reported for $BiSrCaCu_2O_x$ [8]. High-$T_c$ cuprate superconductors have two common characteristics in their crystal structure: the structure contains a $CuO_2$ plane as a structural unit, and superconductivity appears when a carrier is doped in the insulating parent phase with AFM. Although various new superconductors have been reported within these frameworks, no update of $T_c$ has been reported for cuprates since 1993 [9]. A general consensus that the dominant mechanism for cuprate superconductors is not a conventional phonon-mediated mechanism has been reached. The superconducting phase has orthorhombic symmetry for Y- and Bi-systems, and the coherent length corresponding to interacting separation between two electrons forming a Cooper pair is very short (~2 nm). In addition, high-$T_c$ cuprates exhibit large anisotropy in their superconducting properties. Thus, fabrication of superconducting wire/tape requires three-dimensionally controlled crystallites, resulting in high cost because of the complicated fabrication processes.

In 2001, a new high-$T_c$ intermetallic superconductor $MgB_2$ with $T_c$ = 39 K was reported by a group led by Jun Akimitsu [10]. This material has a layered structure similar to high-$T_c$ cuprates but does not exhibit magnetism. Studies on the isotopic effect and electronic heat capacity clarified that the dominant mechanism followed the conventional BCS-type mechanism [11]. The discovery of high $T_c$ in $MgB_2$ with a simple crystal structure and chemical composition was surprising because this material is the first metal-based superconductor beyond the 'BCS wall' [12]. The superconductivity appears without any carrier doping, and the conventional power-in-tube (PIT) method can be used to fabricate wires similar to the process used for other metal alloy superconductors. A technical disadvantage of this system is the insufficient upper critical magnetic field, which is comparable to that of A15-type superconductors, because the primary application of superconducting wires is magnets to generate a strong magnetic field. The maximum $T_c$ has almost been unchanged because of its degradation upon doping with any type of impurity (except a slight increase by B isotope replacement).

IBSCs were discovered by a group led by Hideo Hosono [13, 14] in the course of exploration of magnetic semiconductors as an extension of research on transparent p-type semiconductors. These researchers studied $LaT_MPnO$ ($T_M$ = 3d transition metal, $Pn$ = pnictogen), which has the same crystal structure as LaCuOCh ($Ch$ = chalcogen) composed of an alternate stack of $(CuCh)^-$ and $(LaO)^+$ layers. LaCuOCh is a wide-gap p-type semiconductor that these researchers cultivated. The first IBSC was discovered in 2006 for LaFePO; however, the $T_c$ remained as low as ~4 K [13]. High-$T_c$ materials were subsequently discovered for $LaFeAsO_{1-x}F_x$ with $T_c$ =26 K ($x$ = 0.08) in February 2008 [14], intervening the discovery of LaNiPO with $T_c$ = 3 K in 2007 [15]. Then, a $T_c$ of 43 K was reached under a high pressure of 4 GPa, exceeding that of $MgB_2$ (39 K), in April 2008 [16]. It was widely believed that elements with a large magnetic moment were harmful to the emergence of superconductivity because the magnetism arising from the static ordering of magnetic moments competes with superconductivity, which is needed for the formation of Cooper pairs (dynamic pairing of two conducting electrons with opposite spin). Because iron and nickel are representative magnetic elements, these discoveries were accepted with surprise by the condensed matter physics community, and extensive studies immediately began in China, the US, and Europe. The rapid research progress was described in the proceedings of the first international conference on IBSCs [17] held in Tokyo in 2008 and in the first special issue on IBSCs [18].

A vast number of papers on IBSCs have been published since then. Readers are encouraged to refer to the comprehensive reviews and monographs listed in references [19-23]. The current review describes the recent progress in IBSCs from the viewpoint of materials science and technical applications.

2. Materials

2.1. Parent materials

Since the first paper reporting $T_c$ = 26 K in LaFeAsO$_{1-x}$F$_x$, several tens of superconducting layered iron pnictides and chalcogenides have been reported. These materials contain a common building block of a square lattice of Fe$^{2+}$ ions with tetrahedral coordination with $Pn$ (P and/or As) or $Ch$ ions. To date, a lot of parent compounds are known, and each crystal structure can be derived from the insertion of ions and/or building blocks between the Fe$Pn$($Ch$) layers [20]. Figure 2 shows the structures of the parent materials of IBSCs. Because the Fermi level of each parent compound is primarily governed by five Fe 3d orbitals, iron plays a primary role in the superconductivity. This observation is in sharp contrast to that in cuprate superconductors, in which only one 3d orbital is associated with the Fermi level. In addition, IBSCs have tetragonal symmetry in the superconducting phase, are the Pauli paramagnetic metals in the normal state, and undergo crystallographic/magnetic transition from the tetragonal to orthorhombic or monoclinic phase with AFM at low temperatures. The exceptions are 11- and 111-type compounds exhibiting the Pauli paramagnetism and 245 compounds exhibiting AFM insulating properties.

2.2. Doping to induce superconductivity

Superconductivity emerges when AFM disappears or is diminished by carrier doping, structural modification under external pressure, or chemical pressure via isovalent substitution. In any case, the parent materials are metals with itinerant carriers. Thus, in most cases, removing the magnetism is an experimental step needed for the emergence of superconductivity.

2.2.1. Aliovalent doping

The first high-$T_c$ IBSC was discovered through partial replacement of F$^-$ ions at oxygen sites in La-1111 compounds [14]. The 1111-type compounds consist of a metallic conducting FeAs layer sandwiched by insulating LaO layers. When the O$^{2-}$ site is replaced with an F$^-$ ion, the generated electron is transferred to the FeAs layer because of the energy offset between the layers. Figure 3 presents a schematic phase diagram of the 1111- and 122-type systems. For the 1111 system, $T_c$ appears when the AFM disappears. However, the AFM and superconductivity coexist in the 122 system, and the optimal $T_c$ appears to be achieved at a doping level at which the Néel temperature ($T_N$) reaches 0 K, suggesting the close relationship between the optimal $T_c$ and quantum criticality (electronic phase transition at 0 K). Electron doping of *Re*-1111 compounds (where *Re* = rare-earth metal) via this substitution was very successful; i.e., the optimal $T_c$ from 26 K to 55 K was achieved with the use of Sm (instead of La) with smaller *Re* ionic radius [24, 25].

However, no experimental data on the shape and width of the completely closed $T_c$-dome in the 1111 system with the highest $T_c$ were obtained until 2011 because the electron-doping level was insufficient to observe the over-doped region. The insufficient electron-doping level was attributed to the poor solubility of F⁻ ions at the oxygen sites (approximately 10–15%), resulting from the preferential precipitation of the stable *Re*OF impurity phase. This obstacle was overcome using hydride ions (H⁻) in place of F⁻ [26]. Hydrogen is the simplest bipolar element and can take +1 and −1 charge states depending on its local environment [27]. The ionic radius of H⁻ (110 pm) is also fairly similar to that of F⁻ (133 pm) or $O^{2-}$ (140 pm). H⁻-substituted *Re*-1111 compounds, $Re\text{FeAsO}_{1-x}\text{H}_x$, were successfully synthesized with the aid of high pressure and an anvil cell modified for this synthesis. The synthesis was based on the idea that a H⁻-substituted state would be more stable than an oxygen vacancy state in the charge blocking layer *Re*O with fluorite structure (an oxygen ion occupies a tetrahedral site) [27, 28]. Based on this idea, the mixture of starting materials was heated with a solid hydrogen source such as $CaH_2$, which releases $H_2$ gas at high temperatures under 2 GPa. Figure 4 presents electronic phase diagrams of $Re\text{FeAsO}_{1-x}\text{H}_x$ with different *Re* (La, Ce, Sm, and Gd) [29], which highlight three new findings. First, La-1111 has a two-dome structure. The first dome is the same as that previously reported for $\text{LaFeAsO}_{1-x}\text{F}_x$; however, the second dome unveiled by H doping has a higher optimal $T_c$ (36 K) and larger width. The temperature dependence of electrical resistivity at the normal state (150 K > $T$ > $T_c$) directly above $T_c$ follows $T^2$ (Fermi liquid like) for the first dome but $T^1$ (non-Fermi liquid like) for the second dome. The double-dome structure is not unique to the La-1111 system and has also been observed for chemical compositions with ~30 K > $T_c$ in $\text{SmFeAs}_{1-y}\text{P}_y\text{O}_{1-x}\text{H}_x$ [30], as shown in Fig. 5. Second, although $T_c$ has a single dome for other *Re* systems, its range is much wider than that reported in the F-substituted case for the other *Re* systems. Third, the optimal doping level decreased with decreasing *Re* ion size. The two $T_c$-dome structures in $\text{LaFeAsO}_{1-x}\text{H}_x$ became a single dome under high pressure, and the optimal $T_c$ of 52 K was achieved in the valley at ambient pressure [31]. These observations suggest that when the $T_c$-double-dome structure is transformed into a single dome with a wide width, the optimal $T_c$ > 50 K appears in the *Re*-1111 system regardless of *Re*.

The two-dome structure in the La-1111 system is considered to be derived from the two types of parent compounds with AFM ordering. Both parents are AFM metals; however, their magnetic moments and spin arrangement differ distinctly. Recently, the presence of two parent phases at the extreme edge of the superconducting region was observed for the Sm-1111 system, for which the $T_c$ has a single-dome structure with an optimal $T_c$ > 50 K [32] (see Fig. 4c). These results imply that the high $T_c$ in the 1111 system is realized by the cooperation of two types of fluctuation controlling the two AFM parent phases [32].

Hole doping of the 122 system is possible by substitution of an alkaline-earth ion site with an appropriate alkali ion (e.g., K substitution of the Ba site) [33], which will be discussed later. However, the hole-doping effect in the $Re$1111 system remains unclear.

2.2.2. Revisit of electron doping by oxygen vacancies

Another electron-doping method involves the introduction of an oxygen vacancy in the $Re$O layers in $Re$FeAsO$_{1-x}$ [34, 35]. These samples were synthesized by heating a batch of intentionally prepared oxygen-deficient compositions under high pressure. If a vacancy substitutes for the oxygen ion site, two carrier electrons per vacancy should be generated; however, the results presented in Fig. 6a differ greatly from this expectation. In addition, the $T_c$ values of nominally oxygen-deficient REFeAsO$_{1-x}$ samples agree well with those of hydrogen-substituted ones when plotted against their $a$-axis lattice dimension [28]. To clarify this contradiction, Muraba et al. examined the preferred electron-dopant species at oxygen sites in $Re$FeAsO by changing the atmosphere (H$_2$, H$_2$O, or H$_2$- & H$_2$O-free), ensuring a high-pressure cell assembly to prevent external contamination [36]. The following observations were made: (1) The samples synthesized under an H$_2$ or H$_2$O atmosphere in high-pressure synthesis were $Re$FeAsO$_{1-x}$H$_x$, not $Re$FeAsO$_{1-x}$. (2) The samples with the nominal composition $Re$FeAsO$_{1-x}$ synthesized under H$_2$- and H$_2$O-free atmospheres were nearly stoichiometric $Re$FeAsO. These results strongly suggest that the samples of $Re$FeAsO$_{1-x}$ reported thus far are actually $Re$FeAsO$_{1-x}$H$_x$, which are formed by incorporating hydrogen from the atmosphere and/or starting materials. First-principles calculations substantiated that the hydrogen-substituted samples were more stable than the oxygen-vacancy-substituted ones [36]. A similar observation was also recently reported for amorphous oxide semiconductors in which oxygen vacancy is believed to be the dominant defect [37].

2.2.3. Isovalent doping

A unique characteristic of doping IBSCs is isovalent doping. Two typical examples are introduced. One is partial substitution of the Fe$^{2+}$ site by Co$^{2+}$ and the other is replacement of the As$^{3-}$ site with P$^{3-}$. The former example may be understood in terms of electron doping because the Co$^{2+}$ (3d$^7$) has an excess electron compared with Fe$^{2+}$ (3d$^6$) [38]. This finding contrasts sharply with the results of impurity effects in high-$T_c$ cuprates, for which $T_c$ is easily degraded by partial replacement of the Cu$^{2+}$ site. The robustness of $T_c$ to impurities is closely related to the pairing mechanism. This type of substitution is often called 'direct doping' because the $T_M$ such as Co replaces the iron sites where superconductivity emerges. It is natural to consider that the $T_c$ induced by the direct doping is considerably lower than that induced by indirect doping. Figure 7 compares the direct and indirect doping of $T_c$ in the 122 system (also see Fig. 5a for the 1111 system).

Another effective isovalent substitution is observed in the 122 system such as BaFe$_2$(As$_{1-x}$P$_x$)$_2$ [39]. As the $T_N$ of the parent phase is reduced by $x$, the $T_c$ appears and reaches a maximum of ~30 K around $x=0.35$, which appears to correspond to the quantum critical point. The shape of this phase diagram is similar to that obtained by electron doping using Co substitution. The emergence of superconductivity with the similar isovalent substitution of anions directly bonding with iron is observed for FeSe$_{1-x}$Te$_x$ [40]. Because isovalent anion substitution does not generate carriers (unlike Co substitution), it is understood that the anion substitution modifies the local geometry around iron atoms, which in turn leads to weakening of AFM order competing with the emergence of superconductivity. Because the parent materials of IBSCs are metals containing sufficient carriers to induce superconductivity, it is understood that the primary effect of isovalent anion substitution is to weaken the AFM.

### 2.2.4. Doping by intercalation

The parent materials of IBSCs have layered structures. Insertion of ions and/or molecules is possible while maintaining the original Fe*Pn*(*Ch*) layers in some parent materials. Metal–superconductor conversion via this doping has been reported for 11 and 122 compounds. The FeSe intercalates obtained from low-temperature alkali metal and NH$_3$ co-intercalation exhibit higher $T_c$ of 30–46 K than the samples prepared using conventional high-temperature methods [41]. A unique feature of this process is that a small-sized alkali cation such as Li or Na combined with the NH$_2^-$ anion or NH$_3$ molecules can be intercalated into the FeSe layers [42] because the formation of ion intercalates is restricted to large-sized monovalent cations such as Cs and Tl [43] by conventional high-temperature methods.

When SrFe$_2$As$_2$ thin films are placed in an ambient atmosphere, this film is converted into a superconductor accompanying shrinkage of the *c*-axis [44]. Based on an observation that this conversion does not occur in a dry atmosphere, the intercalation of H$_2$O-relevant species into a vacant site in the Sr layers was suggested [44]. Such a conversion is not observed for BaFe$_2$As$_2$ [45] with a vacancy with smaller space than that in SrFe$_2$As$_2$. This finding led to the shift of thin-film research from SrFe$_2$As$_2$ to BaFe$_2$As$_2$, which is less sensitive to ambient atmosphere [46]. Consequently, research on the 122 system has been performed mainly for BaFe$_2$As$_2$ to date. A similar conversion was reported after immersing the parent compounds into polar organic solvents including wines [47]. Notably, it has been reported that strain can induce a similar effect in a bulk single crystal [48].

### 2.3. Correlation between $T_c$ and local structure

It is a general trend that the optimal $T_c$ is higher in the order 1111 > 122 > 11. This result implies that the optimal $T_c$ is enhanced by the interlayer spacing of FeAs layers. However, this view is not valid [20]. Instead, it is now a consensus that the $T_c$ of IBSCs is sensitive to the local geometry of the Fe*Pn*(*Ch*)$_4$ tetrahedron. Lee et al. [49] first reported that the optimal $T_c$ is achieved when the bond angle of *Pn*(*Ch*)–Fe–*Pn*(*Ch*) approaches that of a regular tetrahedron (109.5º). Figure 8 plots most of the data including the non-optimal $T_c$ values for various types of IBSCs. The phenomenological correlation between $T_c$ and the bond angle becomes worse than that between the optimal $T_c$ and the bond angle; however, the tendency still remains. However, data on the 11 system and the first dome in LaFeAsO$_{1-x}$H$_x$ differ greatly from this empirical rule. This discrepancy stems from $T_c$ not being determined only by the local structure of Fe$Pn$($Ch$)$_4$. Kuroki et al. [50] proposed a model that the pnictogen (chalcogen) height ($h$) from the iron plane is a good structural parameter associated with strength of spin fluctuation and that $T_c$ is enhanced by increasing $h$. The correlation between $h$ and $T_c$ is comparable to that between $T_c$ and the bond angle around Fe.

### 2.4. Advantageous properties for application of IBSCs

Table 1 compares important properties associated with applications of three representative high-$T_c$ superconductors: IBSCs, MgB$_2$, and cuprates (Y- and Bi-systems). The maximum $T_c$ of the cuprates is the highest among them due mainly to strong electron correlation in AFM Mott-insulating state of parents with much high $T_N$ (> room temperature). The $T_c$ of IBSCs is the next class to that of cuprates. The temperature range is not applicable to liquid nitrogen (boiling point: 77 K) like cuprates, whereas we can expect the application of IBSCs in the temperature range of refrigerators (~10 K) and liquid hydrogen (20 K). The unique characteristic of IBSCs is the robustness of $T_c$ to impurity (i.e., doping), which enables us to select various doping methods to induce their high-$T_c$ superconductivity. The upper critical magnetic fields ($H_{c2}$) are quite high well over 50 T, which are higher than those of MgB$_2$ and conventional metallic superconductors such as Nb-Ti (15 T) and A15-type Nb$_3$Sn (29 T). Thus, one of the application targets of IBSCs is expected to be high-magnetic-field magnets. The more important property for practical application for magnets is irreversibility field ($H_{irr}$) because it is the maximum field when critical current density $J_c$ becomes zero. This value is also higher than that of MgB$_2$ in the same temperature range, further expecting future magnetic field application. The anisotropy factors $\gamma$ (= ($m_c$ / $m_{ab}$)$^{1/2}$ = $\xi_{ab}$ / $\xi_c$ = $H_{c2}^{//ab}$ / $H_{c2}^{//c}$, where $m$ and $\xi$ are effective mass and coherent length, respectively) of the IBSCs are comparable to that of MgB$_2$ and quite smaller than those of cuprates. This small $\gamma$, high crystallographic symmetry, and large critical grain boundary (GB) angle ($\theta_c$) for $J_c$ indicate that the IBSCs are appropriate for fabrication of superconducting wires, tapes, and coated conductors' application because high and three dimensional crystallographic orientation is not necessary rather than cuprates. Therefore, it is expected that IBSCs are applicable to wires, tapes, and coated conductors for high magnetic fields as will be discuss in the following sections 3 – 5.

### 3. Bulk magnets
### 3.1. Introduction to superconducting bulk magnets

The trapped field in a superconducting bulk magnet is a direct manifestation of quantum phenomena on the macroscopic scale. Magnetic levitation by perfect diamagnetism or magnetic vortex pinning is one of the most famous demonstrations of high-$T_c$ superconductivity. Superconducting bulk magnets show great potential to largely improve the performance of motors, generators, and analytical instruments by replacing conventional permanent magnets.

The principle of the superconducting bulk magnet is illustrated in Figure 9. By magnetizing below its $T_c$, the entire bulk material acts as a compact Tesla class magnet because of the presence of remotely induced macroscopic circulating supercurrents (Fig. 9). When a magnetized bulk is kept cold, it will behave like a permanent magnet because the induced persistent current does not decay. Another interpretation of the origin of trapped fields is based on quantum vortices, which are compacted and pinned in vortex pinning centers. Generally, the maximum trapped field $B_T$ of a cylindrical superconducting bulk can be expressed using the Bean model:

$$B_T = A\mu_0 J_c^{bulk} r,$$

where $A$ is a geometrical factor (equal to 1 for a cylinder of infinite length), $\mu_0$ is the permeability of vacuum, $J_c^{bulk}$ is the bulk or globally circulating critical current density, and $r$ is the radius of the sample. In analogy to a solenoid electromagnet, the trapped field of a bulk magnet is roughly proportional to $J_c$ and the size (current cross-section). Because of their high current densities, which are more than 100 times that possible in Cu, and size-dependent nature, very high fields (>10 T) are expected in bulk superconducting magnets, substantially higher than those of conventional spin-based ferromagnets, for which the field strength is limited by the intrinsic remanent magnetization ($B_r$), typically less than 2 T. These unique features, *i.e.*, compactness, lightweightness, and very high field, make superconducting bulk magnets interesting for novel permanent magnet applications.

Conventional metallic superconductors, such as Nb–Ti and Nb$_3$Sn, are used in thermally stabilized multifilament wire forms rather than in large bulk forms because their maximum performance is expected for operations at liquid-helium temperature, where thermal instability becomes an issue. After the discovery of cuprate high-temperature superconductors (HTSs), extensive research and development on YBCO ($Re$Ba$_2$Cu$_3$O$_{7-\delta}$) large bulks have been performed. The top seed melt growth (TSMG) process using a seed crystal as the nucleus was developed to obtain quasi-single-crystal bulks with sizes of several centimeters. Our understandings of the material processing, magnetic vortex pinning, and mechanical characteristics have greatly advanced[51-53]. As a result, the trapped magnetic field strength has been largely improved. The remarkable achievement is trapping a very high magnetic field of 17.6 T in 1-inch diameter GdBCO bulks with shrink-fit stainless steel[54]. Another innovation in this area is *pseudo bulk*, in which thin-film-coated conductors cut into square shapes are stacked and magnetized to trap magnetic fields[55,56].

Another research trend is the development of larger bulks using a process different from TSMG. MgB$_2$ discovered in 2001 has 40-K-class high $T_c$[10] and thus can be operated at medium temperatures (10–20 K) without requiring liquid helium. It has been shown that *polycrystalline* bulk MgB$_2$, which can be synthesized using simple ceramic processing, can trap a magnetic field of several tesla. Moreover, the trapped magnetic field in bulk MgB$_2$ exhibits excellent spatial and temporal uniformity[57]. The mechanism by which the high, uniform, and stable field in bulk MgB$_2$ develops is unique compared with that observed in HTS materials and is attributed to grain boundaries acting as homogeneous vortex pinning sites rather than current-blocking defects. Being simple binary compounds without weak-link natures[58], large-sized polycrystalline MgB$_2$ bulks can be produced on an industrial scale. Efforts have been made to increase their trapped magnetic field strength by improving their processing [59-62].

### 3.2. Processing and properties of IBSC polycrystalline bulks

The IBSC discovered in 2008[14] exhibits 60-K-class high $T_c$. Shortly after its discovery, the IBSC was observed to exhibit a very high upper critical field $H_{c2}$, exceeding 100 T in 1111[63-70] and 50 T in 122[71-76], with relatively small electromagnetic anisotropy and thus attracted considerable attention for high-magnetic-field applications[77,78]. The remarkable feature of IBSCs is that high-temperature superconductivity (higher $T_c$, larger anisotropy, stronger thermal fluctuation) [64] and rather conventional metallic superconductivity (lower $T_c$, smaller anisotropy, weaker thermal fluctuation)[74,79] are mixed in various compounds sharing the tetragonal lattice of iron. This feature yields two distinct directions for material developments: towards untextured polycrystalline IBSC materials with random crystal orientation (bulk or round wire, like MgB$_2$) or textured quasi-single-crystalline IBSC materials (epitaxial thin film or textured flat tape, similar to the cuprates). In this section on bulk magnets, we focus on the former direction.

IBSC polycrystalline materials are mainly synthesized using a solid-state reaction method (a ceramic process) in which powdered raw materials are weighed and mixed to form the designed composition and then synthesized by heat treatment after cold working. Some elements with high chemical reactivity and/or high equilibrium vapor pressure (alkali, alkaline-earth, rare-earth, chalcogen, and pnictogen metals) require special precautions; the raw materials may be sealed in metal foil/container or vacuum silica-glass tube or processed under high pressure.

Figure 10 presents examples of the micro- and nanostructure of IBSC polycrystalline bulk samples prepared using the ambient pressure processing described in ref [80]. Figure 10a displays an EBSD-IPF crystal orientation map for the polished surface of a Co-doped Ba122 (Ba122:Co) polycrystalline bulk. Generally, 122 polycrystalline bulk has a structure in which crystals with submicron to tens of microns sizes with random orientation are connected. In addition, voids, cracks, and impurity phases may exist. Figures 10b and c present examples of the grain boundary structure of Ba122:Co. A clean grain boundary is observed in Fig. 10b, whereas an amorphous oxide phase with a thickness of 3–5 nm can be observed at a triple point and grain boundaries in Fig. 10c. The synthesis of high-purity material is one of the issues for IBSC polycrystalline materials[65,81,82] because structural defects and impurities in microstructures (extrinsic ones)[83,84] as well as intrinsic weak links [85-89] behave as transport-current-limiting factors.

Early studies revealed that the transport critical current density ($J_c^{global}$) is suppressed compared with the local critical current density ($J_c^{local}$) in 1111 polycrystalline materials in analogy to cuprates[23,79,81,83,90-94]. When an IBSC polycrystalline bulk is exposed to a magnetic field, the magnetic vortex preferentially penetrates through the grain boundaries rather than into the grains, making it electromagnetically granular. Experiments on the grain boundary misorientation angle ($\theta_{GB}$) dependence of intergranular transport $J_c^{BGB}$ with bicrystal thin films [85-89] explicitly revealed the microscopic origin of the granularity. Katase *et al.* reported that the critical current density through bicrystal grain boundaries ($J_c^{BGB}$) remained high and nearly constant up to a critical angle $\theta_c$ of ~9° for Ba122:Co, which is substantially larger than the $\theta_c$ of ~5° for YBCO, and that even at $\theta_{GB}>\theta_c$, the decay of $J_c^{BGB}$ was much slower than that of YBCO[86]. Thanks to the high $\theta_c$ and low anisotropy, high $J_c^{global}$ values have been achieved in polycrystalline bulks and wires by tuning the processing conditions in 1111 [82,95-97], 122 [80,98-101], and 11 [102-104]; by chemical doping in 1111 [105,106], 122 [107,108], and 11[109]; and by irradiation [110,111] (details can be found in the wire section).

### 3.3. IBSC bulk magnet

Recently, an IBSC bulk magnet capable of providing a powerful magnetic field was demonstrated [112]. Trapped magnetic fields of 1 T at 5 K and 0.5 T at 20 K were obtained in a compact K-doped Ba122 (Ba122:K) bulk with a diameter of 1 cm and height of 2 cm. To synthesize the K-doped Ba122 bulks, elemental Ba, K, Fe, and As powders were reacted together using a mechanochemical reaction[100,101] followed by sintering in a hot isostatic press (HIP) at 600 °C. After bulk synthesis and subsequent re-milling, the Ba122 powder was pressed into pellets and then further densified in a cold isostatic press (CIP). These densified pellets were then wrapped with Ag foil and inserted into a carefully machined steel tube. After welding, the tubes were swaged and CIPped to further shape and densify them, reducing the diameter of the samples by ~10 %. Finally, the samples were sintered at 600°C in the HIP. After the heat treatment, the steel tubes were sliced to reveal the pellet surfaces. The appearance of the resulting Ba122 bulks is shown in Fig. 11. Room-temperature Vickers hardness (HV) tests on the sample revealed an average HV of ~3.5 GPa. Moreover, a fracture toughness of ~2.35 MPa m$^{0.5}$ was calculated from the length of micro-cracks propagating from the corners of the microindentations. The high fracture toughness is attributed to its dense, nano-polycrystalline nature and exceeds those of single-crystal Mn-doped Ba122 [113] and TSMG-processed bulk YBCO [114], suggesting that Ba122:K polycrystalline bulk is interesting for trapped-field applications from the viewpoint of mechanical properties.

Figure 12 shows the trapped field measured by Hall sensors placed on the bottom surface (H1) and between (H2) the stack of Ba122 bulk magnets with ~10 mm in diameter and ~18.4 mm in total thickness. At 5 K, the bulk stack trapped 0.68 T at the center of the outer surface (H1) and 1.02 T on the cylinder axis between the bulks (H2). The trapped field decreased with increasing temperature and vanished at $T_c$ ~33 K. Notably, approximately 50% of the maximum trapped magnetic field (at 5 K) was obtained at ~20 K, which is relatively easily reached by a compact cryocooler. The average macroscopic current density at 5 K was estimated to be ~50 kAcm$^{-2}$ using the Biot–Savart approximation, the total thickness of the magnet stack, and the experimentally determined trapped field of H1. This result matched the $J_c$ value obtained based on local magnetization measurements made on small bulk samples at 4.2 K under 0.6 T. This finding suggests that supercurrent uniformly circulates over the bulk (as confirmed by magneto optical imaging[112]), indicating that the homogeneity of the chemical composition/microstructure was satisfactory.

3.4. Prospects for bulk magnets

Figure 13b plots the maximum trapped field as a function of radius for Ba122:K and $MgB_2$ bulks calculated from the $J_c(H)$ data shown in (a). This calculation is for an infinitely long cylinder geometry and accounts for the radial field dependence of $J_c(H)$. The results indicate that Ba122 bulks with larger radius would be capable of trapping higher fields, whereas $MgB_2$ outperforms Ba122 at low fields and small radius. The remarkable advantage of Ba122 over $MgB_2$ is its weak field dependence of $J_c(H)$ because of its much higher $H_{c2}$ over 50 T at 20 K [115]. Given that large-size bulk IBSC magnets with a polycrystalline microstructure would be easily fabricated using conventional ceramic processes, IBSC may offer unique advantages for large-diameter, high-field magnets that cannot be supplied by either $Re$BCO or $MgB_2$. Such large IBSC bulks would be useful for magnetic levitation, in energy storage applications, and in compact magnetic resonance devices.

4. Thin films and devices

The important six parent phases shown in Figure 2c have provided a wide platform for research on superconducting thin films and devices. Superior superconducting properties such as a high maximum $T_c \approx 55$ K, high upper critical magnetic fields > 50 T, lower anisotropy factors in the superconducting properties than those of cuprates (the 122 compounds have particularly low anisotropy among IBSCs), and an advantageous grain boundary nature [86], appear to be appropriate for wire, tape, and coated conductor applications for ultrahigh-performance magnets.

Since the discovery of IBSCs, several reviews focusing on thin films and devices have been published [20,77,116-123]. In this section, we briefly review the progress and current status of the growth and performance of thin films as well as the device fabrication of IBSCs.

4.1. 1111 thin films

The first 1111-type epitaxial films were grown by pulsed laser deposition (PLD), in which the excitation source of laser ablation was replaced from the usual ultraviolet excimer laser to the second harmonics ($\lambda$ = 532 nm; i.e., visible light) of a Nd:YAG laser to overcome the difficulty of the as-grown phase formation by PLD [124]. However, no La1111 epitaxial films exhibited superconducting transitions mainly because of a lack of fluorine dopant in the epitaxial films. The origin of this difficulty in the phase formation via PLD remains unclear. During the initial research on 1111 film growth, the formation of the 1111 phase and incorporation of the fluorine dopant in thin films were the largest issues. Then, the fabrication of La1111 thin films via a two-step *ex situ* growth method, in which PLD at room temperature for film deposition was combined with post-deposition thermal annealing for crystallization, was reported. $T_c^{onset}$ of the La1111 film was 11 K [125]. A similar difficulty with fluorine incorporation in the 1111 films was also reported for fabrication using molecular beam epitaxy (MBE) [126]. Then, La1111 films that exhibited clear superconducting transitions at $T_c^{onset}$ = 28 K and $T_c^{zero} \approx 20$ K were reported by reducing the oxygen partial pressure during post-deposition thermal annealing in the two-step *ex situ* process [127,128]. For MBE growth, the growth time strongly affected the superconducting properties of Nd1111 films [129]. Epitaxial F-doped Nd1111 films were obtained by optimizing the MBE growth time (a longer time was better for obtaining superconducting Nd1111 films). The $T_c$ values of the Nd1111 films were $T_c^{onset}$ = 48 K and $T_c^{zero}$ = 42 K. These findings imply that the fluorine dopant can be effectively introduced when a longer MBE growth time is employed.

Then, several new F-doping techniques for MBE growth were developed; for example, gallium getter [130]; diffusion from an overlayer of $SmF_3$ [131]; and co-evaporation of $SmF_3$ [132], $FeF_2$ [133], or $FeF_3$ [134]. The current maximum $T_c$ reported for Sm1111 thin films grown by MBE is the same as that of bulk samples ($T_c \approx 56$ K). MBE may be the most effective method for obtaining 1111 thin films because each element source and flux rate can be controlled independently. Very recently, an *in situ* F-doping method has also been reported for PLD [135]. In this case, F diffusion in the Sm1111 films from $CaF_2$ substrates was effectively used during high-temperature growth using Nd:YAG laser PLD. The reported maximum $T_c^{onset}$ was ~40 K. Precise optimization of the Nd:YAG PLD growth conditions contributed to this *in situ* PLD growth of Sm1111 films. In addition, a $J_c$ of ~0.1 MA/cm$^2$ at 4.2 K and scaling anisotropy were reported in [128] for the two-step growth of La1111 films. A higher self-field $J_c$ of > 1 MA/cm$^2$ and in-field $J_c$ of > $10^5$ A/cm$^2$ at 4.2 K under ultrahigh fields such as > 20 T are demonstrated for MBE-grown Sm1111 [136] (Fig. 14) and Nd1111 films [137].

Because of the difficulty of film growth of the 1111 phase, the number of research groups that can successfully grow 1111 films is still limited even though 10 years have been passed since the first report on superconductivity of the 1111 phase. However, we expect that an effective vortex pinning center for 1111 films would further enhance their critical current properties and improve their anisotropic properties for future application.

4.2. 122 thin films

A unique doping method, the partial substitution of the Fe site with Co (i.e., 'direct doping'), was reported to induce superconductivity in 1111 and 122 compounds. The effectiveness of the direct doping is the superior characteristic of IBSCs because generally, superconductivity is severely degraded by disturbances to substructures controlling the Fermi surface. This finding has also contributed to the early realization of thin films and devices using 122-type IBSCs because Co possesses a low vapor pressure and is more easily incorporated in thin films than other dopants with high vapor pressures such as F for 1111 phases and K for 122 phases. In particular, successful incorporation of alkali metals such as K in the films usually requires special techniques such as post-deposition thermal annealing in tightly closed atmospheres because of its extremely high vapor pressures, and K-doped thin film samples are unstable in air. Therefore, K-doped films are not appropriate for high-performance thin-film growth, although K-doped Ba122 has the highest $T_c$ within the Ba122 family. Consequently, direct Co-doping led to the rapid realization of high-quality and high-$J_c$ thin films of Co-doped Ba122, especially in the early stage of IBSC research. Recently, Ni-doped Ba122 films exhibiting comparable properties to Ba122:Co have begun to be reported [138, [139]] because Ni is also an effective low-vapor-pressure dopant to the 122 phase, similar to Co.

For the above reasons, Co-doped Sr122 and Ba122 epitaxial films were first demonstrated in 2008 [140] and in 2009 [141], respectively. The Sr122 phase appears to be more sensitive to air exposure than Ba122 [141], and Ca122 growth is difficult using PLD [142] (whereas MBE growth of Ca122 films was recently reported in [143]). Thus, Ba122:Co has been studied most extensively. The strain effect in Ba122:Co epitaxial films was also examined, leading to high $T_c \approx 25$ K on $SrTiO_3$ [144] and approximately 25–28 K on $CaF_2$ [145, 146, 147], which is slightly higher than that of bulk Ba122:Co (~22 K). To grow high-quality Ba122:Co epitaxial films, two types of effective buffer layers were proposed to solve the in-plane lattice mismatch between Ba122:Co and single-crystal substrates: perovskite oxides such as $SrTiO_3$ [148] and Fe metal layers [149]. In both cases, an ultraviolet KrF excimer laser was employed as an excitation laser for PLD. However, for Nd:YAG PLD [142,150], no buffer layer was necessary to obtain high-performance Ba122:Co epitaxial films. Because the difference between usual KrF PLD and Nd:YAG PLD is interesting, four types of pulsed laser wavelengths (i.e., ArF (193 nm), KrF (248 nm), second harmonics (532 nm), and fundamental (1064 nm) of Nd:YAG) were employed for Ba122:Co growth, and the effect of photon energy and critical factors for Ba122:Co growth were examined [151]. It was clarified that the optimal deposition rate, which can be tuned by adjusting the pulse energy, is independent of laser wavelength. The high-quality Ba122:Co film grown at the optimal pulse energy (i.e., the optimum deposition rate) exhibited high $J_c$ over 1 MA/cm$^2$ irrespective of the laser wavelength. The estimated optimum excitation energy density for KrF was 6.7–10 J/cm$^2$, which is substantially higher than that of second harmonics (2.2–3.2 J/cm$^2$) and fundamental (1.3–1.6 J/cm$^2$) of the Nd:YAG laser. Therefore, a high excitation energy is necessary for high-quality Ba122:Co growth when a KrF excimer laser is employed for PLD. The main origin of the high $J_c > 1$ MA/cm$^2$ in the Ba122:Co films is the $c$-axis-oriented pinning centers [152-156]. Recently, the in-field $J_c$ performance under magnetic fields (i.e., isotropic and high $J_c$ under magnetic fields such as 2.6 MA/cm$^2$ at 9 T) has been greatly improved in Ba122:Co films [147]. In addition to naturally formed external pinning centers, some intentional approaches for realizing a high pinning force and isotropic properties were also examined such as proton irradiation [157], oxygen impurity concentration tuning [156,158], and the use of undoped Ba122/Ba122:Co modulated superlattices [159].

In addition to extensive research on Ba122:Co, P-doped Ba122 (Ba122:P) epitaxial films have been actively examined since 2012 [87,160-165] because their maximum $T_c$ (31 K) is higher than that of Ba122:Co. Thus, we can also expect higher $J_c$ performance mainly because of their high $T_c$. Ba122:P epitaxial films have been successfully obtained using PLD and MBE. As expected, a higher $T_c \approx 30$ K than that of Ba122:Co was reported. Mainly because of its high $T_c$, high self-field $J_c$ (maximum $J_c^{self} > 10$ MA/cm$^2$ [87]) has been achieved in Ba122:P films. In addition, an artificial pinning center, $BaZrO_3$ (BZO), is effective for enhancing the $J_c$ performance of Ba122:P, similar to cuprates [162]. Contributing strong $c$-axis pinning, a fairly isotropic $J_c$ is realized in Ba122:P grown by PLD [164]. High-performance in-field $J_c$ values exceeding 0.1 MA/cm$^2$ at 35 T for $H\|ab$ and 18 T for $H\|c$ at 4.2 K were demonstrated for Ba122:P films grown by MBE [165].

Figure 15 compares the in-field $J_c$ performance of an 1111 film and Ba122:Co and Ba122:P films. The in-field properties of Ba122:Co in the early stage in 2010–2012 were not promising. However, the in-field $J_c$ performance of Ba122:P is generally superior to that of Ba122:Co mainly because of its high $T_c$ and effective vortex pinning. Recently, a high-$J_c$ Ba122:Co film with isotropic $J_c$ was demonstrated on CaF$_2$ substrates [147].

### 4.3. 11 thin films

Initially, there were a few reports of 11 films with zero resistivity, implying the difficulty in fabricating superconducting 11 thin films owing to the complicated phase diagram (i.e., precise control of the chemical composition is necessary.). Among the numerous works on 11 films [40,166-174], the highest $T_c$ to date (except those of monolayer 11) is $T_c^{onset} \approx 20$ K or slightly higher [166, 168], which is higher than that of the bulk 11 phase at ambient pressure (~14 K) and comparable to that of Sr/Ba122:Co thin films. Recently, the in-field $J_c$ performance of 11 films has also rapidly improved, being comparable to that of Ba122:Co [171-174]. Figure 16 plots $J_c$ (*H*) values of 11 films on a CaF$_2$ substrate.

### 4.4. Superconducting thin-film devices
#### 4.4.1. Josephson junctions

The thin-film Josephson junctions (JJs) using IBSCs have been examined using Ba122 and 11 thin films mainly because of the difficulty of phase formation of 1111 films. In 2010, thin-film JJs using Ba122:Co bi-crystal grain boundaries (BGBs) were demonstrated using [001]-tilt (La,Sr)(Al,Ta)O$_3$ (LSAT) bicrystal substrates with a high misorientation angle (30°) [175] after optimizing the film growth conditions and achieving high-quality and high-$J_c$ Ba122:Co epitaxial films. For the BGB junctions, the shape of the *I–V* curve at *B* = 0 mT displays resistively shunted-junction (RSJ)-type behavior without hysteresis. The estimated normal-state resistance ($R_N$) and $R_N A$ (*A* is the cross-sectional area of the junction) of the BGB junctions were 0.012 Ω and 3.0 × 10$^{-10}$ Ωcm$^2$, respectively. However, $I_c$ is clearly suppressed by a weak magnetic field of 0.9 mT. The large $I_c$ modulation of 95% indicates that the Josephson current is responsible for most of the supercurrent through the BGB junction. For the entire temperature range, $J_c$ of the BGB junctions (i.e., inter-grain $J_c$) was approximately 20 times smaller than that of the non-BGB junctions (intra-grain $J_c$), implying that BGBs work as weak-link GBs. The $I_c R_N$ product increased almost linearly with decreasing temperature, and the $I_c R_N$ product at 4 K was 55.8 µV. The two orders of magnitude smaller $I_c R_N$ product compared with that of a YBCO BGB junction originates from the metallic nature of Ba122:Co BGB junctions.

A multi-layered (superconductor–normal metal–superconductor) JJ using a Ba122:Co film and PbIn counter electrodes with a 5-nm-thick Au barrier layer was prepared [176]. Depending on the current bias direction, the *I–V* characteristics of the hybrid junction were slightly asymmetric; however, the *I–V* curve had an RSJ-like nonlinear shape with hysteresis. The $I_c R_N$ product and $J_c$ were 18.4 µV and 39 A/cm$^2$ at 4.2 K, respectively. The *I–V* curves under microwave irradiation at frequencies of 10–18 GHz clearly displayed multiple Shapiro steps, confirming the Josephson effect.

Later, an edge-type [177] and some improved BGB or hybrid-type Ba122:Co JJs were demonstrated [178]. All the $I_cR_N$ products, which are much smaller than YBCO BGBs, did not appear to be drastically improved mainly because of the intrinsic metallic nature of IBSCs. The Josephson effect of a nano-bridge fabricated in 11 films was reported in 2013 [179]. The $I_cR_N$ product of the bridge was as large as 6 mV; however, that of 11 BGB junctions [88, 89] was comparable to that of Ba122:Co BGB junctions.

### 4.4.2. SQUID

A dc superconducting quantum interference device (SQUID) composed of a superconducting loop with two BGB junctions in a Ba122:Co epitaxial film was fabricated on a bicrystal substrate (Fig. 17) [180]. Periodic voltage modulation of $\Delta V = 1.4$ μV was observed in the voltage–flux ($V$–$\Phi$) characteristics of the Ba122:Co dc-SQUID measured at 14 K. Furthermore, a flux-locked loop circuit was employed to evaluate the flux noise $S_\Phi^{1/2}$ spectrum of the dc-SQUID. The $S_\Phi^{1/2}$ level of the Ba122:Co dc-SQUID was more than ten times higher than that of YBCO dc-SQUIDs. The operation temperature of this Ba122:Co dc-SQUID must be close to its $T_c$ because $I_c$ rapidly increased with decreasing temperature. In addition, $V_\Phi$ was low because of the low $R_N$, which was attributed to the metallic nature of normal-state Ba122:Co. Consequently, the high measurement temperature and low $V_\Phi$ were responsible for the high noise level. Therefore, similar to JJs, artificial barriers with large junction resistances, such as superconductor–insulator–superconductor junction structures, are necessary to realize high-performance SQUIDs using IBSCs.

### 4.5. Coated conductors

A critical issue in superconducting wires, tapes, and coated conductors of high-$T_c$ cuprates is the grain boundary (GB) issue. The grains of high-$T_c$ cuprates must be highly oriented to prevent the deterioration of $J_c$ across misaligned GBs because $J_c$ strongly depends on the misorientation angle of GBs ($\theta_{GB}$). For example, a fundamental study on the intergrain $J_c$ ($J_c^{BGB}$) for YBCO was conducted using several types of bicrystal substrates. Significantly misaligned adjacent grains cause $J_c^{BGB}$ to decay exponentially as a function of $\theta_{GB}$ from 3° to 40°. Therefore, to produce YBCO superconducting coated conductors with a high $J_c$, well-aligned buffer layers with a small in-plane misalignment of $\Delta\phi \ll 5°$ on polycrystalline metal substrates must be inserted using an ion-beam-assisted deposition (IBAD) technique or rolling-assisted biaxially textured substrates (RABiTS). However, it has been reported that an IBSC Ba122:Co has an advantageous GB nature compared with cuprates. That is, $J_c^{BGB}$ of Ba122:Co has a gentler $\theta_{GB}$ dependence than that of YBCO, as demonstrated using high-$J_c$ Ba122:Co epitaxial films on [001]-tilt LSAT and MgO bicrystal substrates with $\theta_{GB} = 3°–45°$. The deterioration of $J_c^{BGB}$ due to the tilted GBs is negligible at $\theta_{GB}$ lower than $\theta_c = 9°$ [86], which is twice as large as $\theta_c \approx 3°–5°$ for YBCO BGBs. A similar critical angle for $J_c$, $\theta_c = 9°$, was also reported for 11 films [89]. Therefore, the large $\theta_c$ allows a simple and low-cost production process to be used to produce superconducting coated conductors of IBSCs.

The first coated conductors were demonstrated using Ba122:Co films [181 182 183] with IBAD-MgO technical metal-tape substrates mainly because high-$J_c$ Ba122:Co films on single crystals were achieved in the early stage rather than other IBSC films. The IBAD-MgO substrates consisted of a homoepitaxial MgO layer/IBAD-MgO layer/$Y_2O_3$ buffer layer/Hastelloy C276 polycrystalline tape with $\Delta\phi_{MgO}$ = 5.5°–7.3°. Values of self-field $J_c$ of 1.2–3.6 MA/cm$^2$ were obtained at 2 K ($J_c$ remained above 1 MA/cm$^2$ at 10 K) [182]. The in-field $J_c$ of the Ba122:Co films on the IBAD substrates was substantially higher than that for the films on MgO single crystals, most likely due to $c$-axis vortex pinning effects. These results imply that high-$J_c$ coated conductors can be fabricated with Ba122:Co using less-well-textured templates and with large $\phi$, which allows for a simple and low-cost process for high-$J_c$ and high-$H_c$ superconducting tapes.

Then, some 11 coated conductors were demonstrated on IBAD-MgO [184,185] and RABiTS [186]. In case of IBAD, the self-field $J_c$ (on the order of 0.1 MA/cm$^2$) was slightly lower than that of Ba122:Co conductors, whereas the in-field properties appeared to be better than those of Ba122:Co (well over 0.1 MA/cm$^2$ at 9 T). In particular, for RABiTS [186], superior in-field $J_c$ properties were reported by employing a $CeO_2$ buffer layer (0.1 MA/cm$^2$ at 30 T), as shown in Fig. 18.

Only one 1111-type coated conductor has been reported [187]. The $T_c$ (43 K) was the highest among IBSC coated conductors. $J_c^{self}$ was 0.07 MA/cm$^2$ at 5 K, and the in-field properties were poorer than those of Ba122:Co. The authors suggested that the grain boundaries in 1111 reduce $J_c$ significantly compared with that in Ba122:Co and 11 and, hence, that more biaxial texture is necessary for high $J_c$.

Recently, Ba122:P conductors have been demonstrated as high-performance coated conductors [188,189]. A Ba122:P film exhibited higher $J_c$ at 4 K when grown on a poorly aligned (8°) metal-tape substrate than on a well-aligned (4°) substrate even though the crystallinity was poorer (see Fig. 19). The observed strong pinning in the Ba122:P was attributed to the high-density grain boundaries with misorientation angles smaller than the critical angle. This result reveals a distinct advantage of Ba122:P conductors over cuprate-coated conductors because well-aligned metal-tape substrates are not necessary for practical applications of IBSCs. Recently, the in-field transport properties of the Ba122:P conductor on a poorly aligned substrate were examined in detail [189]. The transport $J_c$ of the Ba122:P coated conductor exceeded 0.1 MA/cm$^2$ at 15 T for the main crystallographic directions of the applied field, which is favorable for practical applications. In addition, the conductor exhibited a superior in-field $J_c$ compared with $MgB_2$ and NbTi and comparable in-field $J_c$ as $Nb_3Sn$ above 20 T. Similar poorly aligned metal tape substrates (7.7°) were also applicable for an 11 coated conductor [185]. Therefore, this usefulness is a powerful advantage of 122 and 11 coated conductors for future fabrication processes.

5. Wires and tapes

There are four main types of IBSCs [190]: SmOFeAs ($T_c$= 55 K, 1111 type), Sr/BaFe$_2$As$_2$:K ($T_c$=

38 K, 122 type), LiFeAs ($T_c$ =18 K, 111 type), and FeSe ($T_c$= 8 K, 11 type). Among these IBSCs, the 122 type is the most relevant for applications because of its ultrahigh $H_{c2}$ > 70 T at 20 K [191], low anisotropy (γ < 2 for 122) [73,115], and ease of fabrication. Shortly after the discovery of superconductivity in pnictides, the first prototype wire based on 1111 and 122 types was prepared by researchers of the IEECAS group using a PIT process [192,193]. Later, the Hosono group clarified that the weak-link effect in 122 pnictides was not as heavy as in YBCO because the critical angle that begins to affect supercurrents reaches up to 9° for pnictides [86], which is substantially larger than the 3°–5° values of YBCO [194]. This finding implies that a scalable PIT technique could be applied for the fabrication of pnictide wires and tapes. Since then, tremendous advances towards improving the $J_c$ in high magnetic fields have been made, especially for 122 family pnictide wires [195-197]. Thus far, the transport $J_c$ values are already extremely high, on the order of $10^5$ A/cm$^2$ at 10 T and 4.2 K, surpassing the widely accepted threshold for practical application. More importantly, the milestone work of the world's first 100-m-class 122-type IBSC wire was achieved in August 2016 [198], which demonstrates the great potential for large-scale manufacture. Therefore, 122-based pnictide wires are of particular interest for their potential to create high magnetic fields well beyond the capabilities of NbTi, Nb$_3$Sn, and MgB$_2$ wires [191, 199].

The following sections provide a review of a few key advances that have enabled the development of high-performance IBSC wires and tapes.

5.1. Fabrication of wires and tapes

The PIT method has been widely used for the fabrication of Bi2223 and MgB$_2$ wires with kilometer length. The PIT process is very attractive from the aspect of applications, taking advantage of the low costs and relatively simple deformation techniques. There are two PIT methods: an *in situ* method, in which a powder mixture of raw precursor is used as a starting material, and an *ex situ* method, in which powder of the reacted superconducting phase is used. In the early years of pnictide wire development, the *in situ* PIT method was used; however, the transport $J_c$ was very low because of the presence of a large amount of impurities, cracks, and voids within the superconducting core. To overcome these issues, Qi et al. produced the first *ex situ* Sr-122 wires by filling Fe/Ag tubes with reacted K-doped Sr122 powder and deforming them into fine wires, followed by annealing at 900°C for 20 h [200]. This process leads to fewer impurity phases as well as a high density of the superconducting core for the final wires. The basic PIT process is schematically illustrated in Fig. 20. Over time, the *ex situ* PIT approach was adopted by most research groups and became the most successful approach for fabricating high-performance pnictide wires. To date, remarkably high $J_c$ values have been obtained in *ex situ* IBSC wires and tapes: $J_c$ (4.2 K) >$10^4$–$10^5$ A/cm$^2$.

In Fig. 20, the choices of the metal tube forming the sheath have been reduced to those elements or metals showing little or no reaction with pnictide during the final heat treatment. The use of various sheath materials has been attempted for fabricating metal-clad pnictide wires and tapes, such as Ag [201, 202], Nb [193, 203], Fe [204-206], and Cu [207,208]. Thus far, Ag has been observed to be the most appropriate sheath material for wire fabrication, showing little reaction with the superconducting phase at the optimized temperatures of the final thermal treatments. Ag may also be used in combination with an additional outer sheath made of Fe, Cu, or stainless steel [100,201,209] to reduce costs and enhance the mechanical reinforcement. Recently, Cu and Fe have also appeared as strong alternatives when the final heat treatment time is short [196]. Because Cu is less expensive and has better deformation properties than Ag, it remains an interesting alternative for industrial applications [208].

A distinctly different situation was encountered for both Bi-2212 and 2223 HTS wires, which require the use of a Ag sheath because Ag is the only material that is inert to the BSCCO superconductor and permeable to oxygen at the annealing temperature [210]. Thus, Ag contributed to at least 60–70% of the BSCCO wire cost. For pnictide wires, in addition to Ag, many types of sheaths of Cu, Fe, and Ag-based composites (Ag/Fe, Ag/Cu, Ag/stainless steel) can be employed, suggesting that pnictide wires are expected to be much more cost effective than BSCCO conductors.

### 5.2. Strategies to improve $J_c$

The improvement of $J_c$ of IBSC wires and tapes developed using the PIT method has been a focus of research and development in this field over the past years, with the value increasing rapidly over time, as shown in Fig. 21. To date, the critical current density of 122-based superconducting tapes has exceeded $10^5$ A/cm$^2$ at 4.2 K under 14 T [211], demonstrating their excellent potential for high-field applications. Figure 21 shows that the in-field $J_c$ values of 1111- and 11-type wires are still too low, e.g., more than 2–3 orders of magnitude lower in the high-field region than those of 122 wires, suggesting the large difficulties in wire fabrication for both 1111 and 11 families. The problem for 1111 wires is controlling the fluorine content during sintering [196], whereas it is very difficult to remove excess Fe for 11 wires [212,213].

Early efforts at wire development suffered from impurity phases that wet the grain boundaries, in particular, non-superconducting amorphous layers surrounding the grains, which show significant oxygen enrichment according to electron energy loss spectroscopy studies [214]. These oxygen-rich layers undoubtedly obstructed many grain boundaries, consequently resulting in a current-blocking effect. This fact was also corroborated by the results of Kim et al.[215] who observed increased O content at the grain boundaries, which was partly responsible for the low $J_c$ across grain boundaries in wire samples. Hence, determining how to reduce the amount of grain-boundary glassy layers, as well as the porosity within grains became a key issue. Chemical addition has been extensively studied as a possible solution to enhance the intergrain $J_c$. It was observed that Ag addition effectively suppressed the formation of a glassy phase as well as an amorphous layer surrounding grains, significantly improved the intergrain connectivity, and reduced the porosity of 122 tapes, leading to a three-fold increase in the transport $J_c$ in high fields [216, 202]. In fact, many groups also observed that Ag-rich and As-rich material may strengthen grain connectivity and help to improve the transport $J_c$ [217-219]. Recently, it was reported that Ag addition even has positive effects in enhancing the $T_c$, $H_{c2}$, and pinning energy in FeSe polycrystals [103]. In contrast, Pb addition promotes grain growth and is beneficial for grain connectivity, with a substantial $J_c$ improvement achieved at low fields with the addition of 5 wt% Pb [220]. Sn, Zn, and In additions have shown similar positive effects in 122 tape samples [221]. It was also reported that over-doping of K in 122 polycrystalline samples strongly enhanced $J_c$ [222] because of the increase in lattice dislocations within the grains, resulting in an enhanced pinning force and hence improved $J_c$. This fact was confirmed by recent less-K-doped tape results ($x = 0.3$) [223]; the transport $J_c$ increased at low fields of below 1 T but decreased in the higher-field region. Some researchers have attempted to improve phase purity by optimizing the thermal treatment [224], two-step synthesis [225], or mechanical alloying of the precursor [101,226].

Similarly, in the Sm-1111 system, Sn was observed to be an effective additive, which can reduce the loss of F, depress the impurity phases, and enhance the intergrain coupling for Sm-1111/Fe tapes, resulting in a large transport $J_c$ of $2.2 \times 10^4$ A/cm$^2$ at 4.2 K in a self-field [227]. Pre-sintering of Sn-added powders enables the reduction of the FeAs wetting phase and fill the voids between Sm-1111 grains, yielding improved grain connectivity and transport $J_c$ up to $3.95 \times 10^4$ A/cm$^2$ at 4.2 K, SF [97]. More recently, by further improving the purity of the superconducting core in Sm-1111/Fe tapes, at 4.2 K, the transport $J_c$ values reached $1.8 \times 10^4$ A/cm$^2$ at 0.6 T and $2.9 \times 10^2$ A/cm$^2$ at 10 T, which are the highest $J_c$ values achieved to date for a 1111 wire [228]. Notably, several groups are attempting wire fabrication using new materials. The Sefat group reported the first wire of a non-As Ba(NH$_3$)Fe$_2$Se$_2$ superconductor using the PIT method, although the $J_c$ was not high [229]. Ba(NH$_3$)Fe$_2$Se$_2$ exhibits lower toxicity than As 122-type IBSCs (e.g. Ba$_{0.6}$K$_{0.4}$Fe$_2$As$_2$). Iyo et al. [219] produced PIT-processed (Sr$_{1-x}$Na$_x$)Fe$_2$As$_2$ tapes, which exhibited a transport $J_c$ of ~ $10^4$ A/cm$^2$ at 20 K under a magnetic field of 2.5 T. Recently, Dong et al. [230] prepared BaFe$_{2-x}$Co$_x$As$_2$ superconducting tapes via the PIT method. The transport $J_c$ at 4.2 K and 10 T was $10^3$ A/cm$^2$.

In 2011, it was understood that 122 iron-pnictides also suffer from the weak-link problem but that it is not as severe as in YBCO [86], e.g., the critical GB angle of 9° (at which the intergrain supercurrents start to degrade) is much larger than the 3°–5° angles for YBCO. For practical use of superconducting wires, weak links between superconducting grains should be improved. Studies of the deformation process by the IEECAS group [205] demonstrated that the 122 grains become highly textured in Fe-sheathed Sr-122 flat tapes by rolling with a large reduction ratio, which is a scalable process that can be used to manufacture the kilometers of 122 flat tape that will be needed for practical applications. This process combined with Sn addition results in significant enhancement of both the core density and $c$-axis texturing of the 122 phase. Correspondingly, the grain boundaries with alignment angles below 9° can become a favorable current path, accompanied by a higher $J_c$ across grain boundaries. Because of the improved texture of 122 tapes, the current density was increased by more than an order of magnitude in magnetic fields at 10 T to above $10^4$ A/cm$^2$ at 4.2 K [231, 206]. A more encouraging result for enhancing $J_c$ was achieved by Dong et al. [225] for Ag-clad Ba-122 tapes using a high-quality precursor in addition to the aforementioned rolling technique. At 4.2 K and 10 T, a higher transport $J_c$ of $5.4 \times 10^4$ A/cm$^2$ and an engineering $J_e$ of $1.5 \times 10^4$ A/cm$^2$ were achieved.

Unlike the high-$T_c$ cuprates, which have much larger values of $J_c^{ab}/J_c^c$ anisotropy, the anisotropy of $J_c$ with the applied magnetic field angle for the textured 122 tapes is quite small, less than 1.5, as reported by Awaji et al. [232]. This finding implies that $J_c$ changes less with the angle between the applied magnetic field and tape plane. As with 2223 and YBCO tapes, $J_c$ is highest with the magnetic field parallel to the ab plane. This low anisotropy behavior of the 122 family of materials means that the tape performance is independent of the field direction, which is beneficial for practical applications.

Densification is another dominant factor that determines the $J_c$ performance of PIT pnictide wire, as porosity severely disrupts the current flow between superconducting grains. Higher $J_c$ in a wire requires a high mass density of the superconducting filaments simply because of the enhancement of the current flow path [233]. In well-connected 122 thin films, values >$10^6$ A/cm$^2$ have been reported [164].

A HIP was first adopted by the Florida group to fabricate Ba-122 round wires [100]. They utilized a high pressure of 192 MPa at 600°C for 20 h to achieve a highly dense core and thus large $J_c$ (at 4.2 K) of $10^5$ A/cm$^2$ in self-field and ~$8.5 \times 10^3$ A/cm$^2$ at 10 T. The University of Tokyo group optimized the HIP process to further increase $J_c$ of the round wires at 10 T to exceed $2 \times 10^4$ A/cm$^2$ [234,235]. The results indicate the importance of the densification of the 122-type core for achieving high $J_c$. A later study showed that the field dependence of the $J_c$ of the polycrystal 122 wire could be significantly reduced by reducing the grain size, which resulted in much higher currents at high magnetic fields [236]. However, 122 round wire lacked texture and used a low-temperature heat treatment. Hence, further $J_c$ enhancement is possible by either introducing grain texture or using a higher heat-treatment temperature. The advantage of the HIP process is the ability to fabricate round wire, which is desirable for magnet applications. The application of hydrostatic pressure was also observed to significantly enhance $J_c$ in NaFe$_{0.97}$Co$_{0.03}$As and Ba$_{0.6}$K$_{0.4}$Fe$_2$As$_2$ single crystals [237,238] because both the pinning center number density and pinning force were greatly increased by the hydrostatic pressure and enhanced the pinning.

Efforts to improve the density and texture of the 122 core have focused on using uniaxial pressing techniques, which have been proven to be effective in enhancing $J_c$. The IEECAS group was the first to report cold pressing of Fe-sheathed 122-type wire, which improved the mass density and induced c-axis texture and ultimately higher $J_c$ [239]. The NIMS group soon fabricated 122/Ag tapes by applying a combined process of flat rolling and uniaxial pressing to achieve a denser structure in the 122 filament ($J_c = 2.1 \times 10^4$ A/cm$^2$ at 4.2 K and 10 T) [240]. The same group achieved better enhancement of a repeated rolling and cold-pressing process under a high pressure of ~2 GPa, yielding transport $J_c$ values above $10^5$ A/cm$^2$ at 6 T, still as high as $8.6 \times 10^4$ A/cm$^2$ at 10 T [241]. This result was attributed to a change in the crack structure and the more uniform deformation achieved by pressing rather than rolling. However, cold pressing always results in fatal micro-cracks inside the superconducting core, which cannot be healed by subsequent heat treatment.

A significant breakthrough in enhancing $J_c$ of IBSC wires was achieved by the IEECAS group using a hot-pressing technique [242]. As shown in Fig. 22, the transport $J_c$ increased to above $10^5$ A/cm$^2$ at 10 T and 4.2 K for the first time, reaching the widely accepted value for practical application. For the flat-rolled tapes, $J_c$ reached a maximum value of $3\times10^4$ A/cm$^2$ at 10 T and 4.2 K. A clear improvement of $J_c$ by a factor of 3.4 was achieved at 4.2 K upon pressing at ~30 MPa at 850°C for 30 min. In addition, the hot-pressed 122 samples exhibited superior $J_c$ values compared with MgB$_2$ and NbTi in the region above 10 T. Through temperature optimization, a new record transport $J_c$ of $10^5$ A/cm$^2$ (4.2 K, 14 T) was achieved in 122 tapes hot pressed at 900°C [211]. This value is by far the highest ever reported in published studies. Examination of the microstructure revealed that hot pressing makes the grains more flexible to couple with each other without producing a large number of crushed grains, thus significantly reducing the voids and cracks and leading to a high mass density. In addition, the grain boundaries are clear without any second phases. Electron backscatter diffraction measurements [211] revealed that the dominant orientation of the grains was (00$l$) and that a large proportion of misorientation angles of the GBs were smaller than 9° such that the transport $J_c$ would not be greatly reduced upon encountering these GBs. All these results suggest that the hot-pressed tape exhibits good grain connectivity and is highly textured. The main challenge is determining how to transfer this already achieved high $J_c$ to the large scale required for high-field applications.

In addition to the connectivity and texture, vortex pinning and motion also play important roles in controlling $J_c$. Studies on single crystals of IBSCs indicate that small-sized normal cores [243] or fluctuating magnetic/structural domains [244] are responsible for the vortex pining. Recently, Dong et al. [245] studied the vortex pinning mechanism and dynamics of high-performance Sr-122 tapes fabricated by hot pressing. It was observed that even though grain boundary pinning was dominant in the vortex pinning mechanism, a large density of dislocations can also serve as effective pinning centers. Moreover, the accordance of the experimental data and $\delta l$ pinning curve indicates that the fluctuation of the mean free path is the main pinning source. It is observed that a small vortex relaxation rate exhibits a weak temperature and field dependence, indicating a strong pinning force. As a result, increasing the density of grain boundaries by decreasing the grain size is a feasible approach to further increase the pinning force. However, point pinning is more efficient than surface pinning because of its smaller size, which is comparable to the coherence length. Therefore, artificially introducing point defects using irradiation or nanoparticle inclusion may be another approach to further increase the flux pinning force and enhance the current-carrying ability.

5.3. Practical properties of pnictide wires

In addition to high $J_c$, there are many other performance requirements for conductor applications. Superconducting conductors for large-scale applications must have a multifilamentary architecture, excellent mechanical properties, large engineering current densities $J_e$ of $>10^4$ A/cm$^2$ in magnetic fields, and small anisotropy of $J_c$ with respect to field direction while also being low cost and enabling scalability of fabrication [246].

Toward practical applications of IBSCs, the fabrication of multifilamentary wires and tapes is an indispensable step to reduce the AC loss induced by the eddy currents in the metallic matrix and the interfilamentary coupling currents. Yao et al. fabricated 7-core Ag/Fe composite $Sr_{0.6}K_{0.4}Fe_2As_2$ tapes by restacking round 122/Ag wire in a second Fe tube, drawing, and rolling, resulting in a $J_c$ of $1.4\times10^4$ A/cm$^2$ at 10 T and a very weak magnetic field dependence at high fields [247,248]. To date, hot-pressed 7- and 19-core Sr-122 tapes have possessed the highest $J_c$ values of $6.1\times10^4$ and $3.5\times10^4$ A/cm$^2$ at 4.2 K and 10 T, respectively [242]. Recently, this group has also fabricated 114-filament Sr-122/Ag/Fe conductors with high $J_c$ [248], in which an average cross-sectional filament size smaller than 50 μm was achieved by drawing the conductors into round wires with 2.0-mm diameters, as illustrated in Fig. 23. When the round wires were flat-rolled into tapes, the transport $J_c$ gradually increased with the reduction of tape thickness. Note that these 114-filament samples exhibited a weak field dependence of $J_c$ up to 14 T, even though they had different sizes and shapes, suggesting their promising potential for high-field magnet applications.

For applications in high magnetic field in which high electromagnetic forces are present, the conductor strength and its tolerance to the mechanical load are important issues. Kovac et al. reported the first results on the electromechanical properties of Sr-122/Ag tapes prepared using an *ex situ* PIT process [249]. The critical current of the Sr-122/Ag tapes increased up to ε ~ 0.2% and then started to decrease with further rapid degradation at a strain above 0.25%, quite similar to that of Bi-2212/Ag wires. More recently, the discovery of reversible critical current performance under a large compressive strain of ε = –0.6 % for Sr-122/Ag tapes was reported by Liu et al. [250] using the so-called U-spring setup over a wide range of applied axial strains (from –0.6 % to +0.3 %) at 4.2 K and 10 T. These results demonstrated that the critical current of 122 pnictide wire exhibits less sensitivity under applied compressive strains than that of Nb$_3$Sn, which is one of the workhorse superconductors for applications. In addition, the ac loss and quench behavior of Sr-122/Ag tapes at different temperatures were investigated [251,252], which were more positive than those of LTS and MgB$_2$.

Ag may be also used in combination with an additional outer sheath made of Fe, Cu, and stainless steel to reduce costs and improve the mechanical strength, e.g., the NIMS group produced Ag/stainless steel double-sheathed Ba-122 tapes [209]. After rolling and heat treating the double-sheath architecture, a transport $J_c$ of ~7.7×10$^4$ A/cm$^2$ (4.2 K, 10 T) was maintained. In addition, the IEECAS group fabricated Ag/Fe Sr-122 composite tapes with high $J_c$ [248]. Very recently, copper-sheathed Sr-122 tapes were fabricated with transport $J_c$ of ~3.5×10$^4$ A/cm$^2$ at 10 T and 1.6×10$^4$ A/cm$^2$ at 26 T at 4.2 K, respectively. Furthermore, the engineering $J_e$ value of the samples was over 10$^4$ A/cm$^2$ at 10 T and 4.2 K [253]. This finding was noteworthy because the use of copper as a sheath material is cost effective, results in good mechanical properties, and provides reliable thermal stabilization in a magnet during transients. Zhang et al. [254] soon made Cu-sheathed Sm-1111 tapes using a low-temperature (400°C) hot-pressing technique, attaining a $J_c$ of 2.37×10$^4$ A/cm$^2$ at 4.2 K, SF. These data proved that a Cu sheath could be applied for IBSC wires and enables similar transport $J_c$ as Ag-sheathed tape, which may significantly inspire research on more practically desirable Cu-sheathed pnictide wires for applications.

5.4. Scaling up to the first 100-m-class 122 wire

A high $J_c$ of $10^5$ A/cm$^2$ at 4.2 K has been achieved by several groups, such as the IEECAS, NIMS, Florida, and the University of Tokyo groups. However, practical applications require not only high $J_c$ conductors but also sufficiently long lengths for winding the coils or cabling. In addition, an easy and simple process is needed to balance the high performance with the production cost of the long wires.

Based on experience with the high-performance short samples and subsequently developed multifilamentary wire process described previously, the IEECAS group started research on fabricating long IBSC wires. In 2014, they produced the first 11-m-long $Sr_{1-x}K_xFe_2As_2$/Ag tape by rolling, which is a scalable industrial process to manufacture long wire [196]. They then progressed to fabricating 100-m-class wires, for which high-level homogeneity of the precursor powder and uniformity of the deformation process are indispensable. In August 2016, this rolling process was scaled up to produce the first 115-m-long 7-filament $Sr_{1-x}K_xFe_2As_2$/Ag superconducting tape [198], as shown in Fig. 24. The transport $J_c$ at 4.2 K and 10 T was measured along the length of the 115-m Sr-122 tape. A very uniform $J_c$ distribution was observed throughout the tape; in particular, an average $J_c$ of $1.3\times10^4$ A/cm$^2$ at 10 T was achieved over the 115-m length.

The achievement of fabricating 100-m-class prototype wires was a key breakthrough and is of great significance for promoting practical applications of IBSCs. Assuming that the costs can be kept low, competition with NbTi and Nb$_3$Sn conductors appears probable, whereas competition with either BSCCO-based or YBCO-based HTS conductors will depend on future progress.

6. Perspectives

Although the $T_c$ of IBSCs is rather low compared with that of cuprates, the IBSCs have distinct advantages in terms of their grain boundary nature, low anisotropy, and high crystallographic symmetry of the superconducting phases (see Table 1). These advantages make it possible to apply the standard processing for alloy superconductors, the PIT method, to fabricate wires and tapes, as described in Section 5. Recent success in the fabrication of >100-m-long wire with a practical $J_c$ represents a milestone, casting a bright light for the future. Figure 25 highlights the plausible application space of the $T_c$–magnetic field relation. A major potential application area of superconducting wires is for high magnetic fields. Robustness to higher magnetic fields is intrinsic to the nature of IBSCs. Recent results on the magnetic-field dependence of $J_c$ of wires and tapes of IBSCs have experimentally demonstrated their excellent robustness to higher magnetic fields. As the performance of refrigeration systems has greatly advanced, the operation of superconductors at 20–30 K is possible without liquid He. This situation provides an opportunity for the application of current IBSCs to higher magnetic fields. The realization of a compact NMR instrument using a liquid-He-free bulk superconducting magnet would also be a convenient landmark.

Although research progress in IBSCs has been rapid to date, the time since their discovery has only been 10 years. Before 2008, Fe-based high-$T_c$ superconductors were beyond the imagination of experts. Extensive research on IBSCs in the past decade has clarified the rich variety of superconducting materials and sophisticated mechanism arising from the multi-orbital nature of IBSCs [255]. The recent discovery of the drastic increase in $T_c$ in single-layer FeSe [256] suggests that many surprises remain unveiled for IBSCs. The discovery of a new superconductor with higher $T_c$ and practical properties appropriate for application is desired. In most cases, high-$T_c$ materials have less desirable properties for application; therefore, metal alloy superconductors are still used for practical application. It is expected that IBSCs with moderate $T_c$ and properties suitable for fabrication would open a new application field for superconductors. The effectiveness of various modes of doping is one of the outstanding characteristics of IBSCs, which results in a rich variety of superconducting materials. We expect that the iron age will become a reality for superconductors with new superconducting materials and improvement of current materials. *Iron is still hot!*

Acknowledgments
The work in Japan was supported by the Ministry of Education, Culture, Sports, Science, and Technology (MEXT) through the Element Strategy Initiative to Form Core Research Center. H. Ho. thanks Dr. Soshi Iimura for help in drawing figures. H. Hi. was also supported by the Japan Society for the Promotion of Science (JSPS) through a Grant-in-Aid for Scientific Research on Innovative Areas "Nano Informatics" (Grant No. 25106007), a Grant-in-Aid for Scientific Research (A) (Grant No. 17H01318) from JSPS, and Support for Tokyotech Advanced Research (STAR). A. Y. was also supported by a Grant-in-Aid for Young Scientists (A) (Grant No. 15H05519) from JSPS. The work in China was supported by the National Natural Science Foundation of China (No. 51320105015), the Beijing Municipal Science and Technology Commission (No. Z171100002017006), and the Bureau of Frontier Sciences and Education, Chinese Academy of Sciences (QYZDJ-SSW-JSC026).

Table 1. Comparison of three representative high-$T_c$ superconductors

| | IBSCs | MgB$_2$ | Cuprates |
|---|---|---|---|
| Parent material | AFM semimetal ($T_N$~150K) | Pauli paramagnetic metal | AFM Mott insulator ($T_N$~400K) |
| Fermi level | Fe 3d 5-orbitals | B 2p 2-orbitals | Cu 3d single orbital |
| Maximum $T_c$ (K) | 55 (for 1111 type), 38 (for 122 type) | 39 | 93 (YBCO), 110 (Bi2223) |
| Impurity | Robust | Sensitive | Sensitive |
| SC gap symmetry | Extended s-wave | s-wave | d-wave |
| Upper critical field at 0 K, $H_{c2}(0)$ (T) | 100 – 200 (for 1111 type) 50 – 100 (for 122 type) ~50 (for 11 type) | 40 | > 100 |
| Irreversibility field, $H_{irr}$ (T) | > 50 (4 K) > 15 (20 K) | > 25 (4 K) > 10 (20 K) | > 10 (77 K, YBCO) |
| Anisotropy, $\gamma$ | 4 – 5 (for 1111 type) 1 – 2 (for 122- and 11-types) | 2 | 5–7 (YBCO), 50–90 (Bi-system) |
| Crystallographic symmetry in SC state | Tetragonal | Hexagonal | Orthorhombic (Y- and Bi-systems) |
| Critical GB angle, $\theta_c$ (deg.) | 8 – 9 | No data | 3 – 5 (YBCO) |
| Advantage | High $H_{c2}(0)$, Easy fabrication | Easy fabrication | High $T_c$ and $H_{c2}(0)$ |
| Disadvantage | Toxicity | Low $H_{c2}(0)$ | High cost due to 3D alignment of crystallites |

Figures

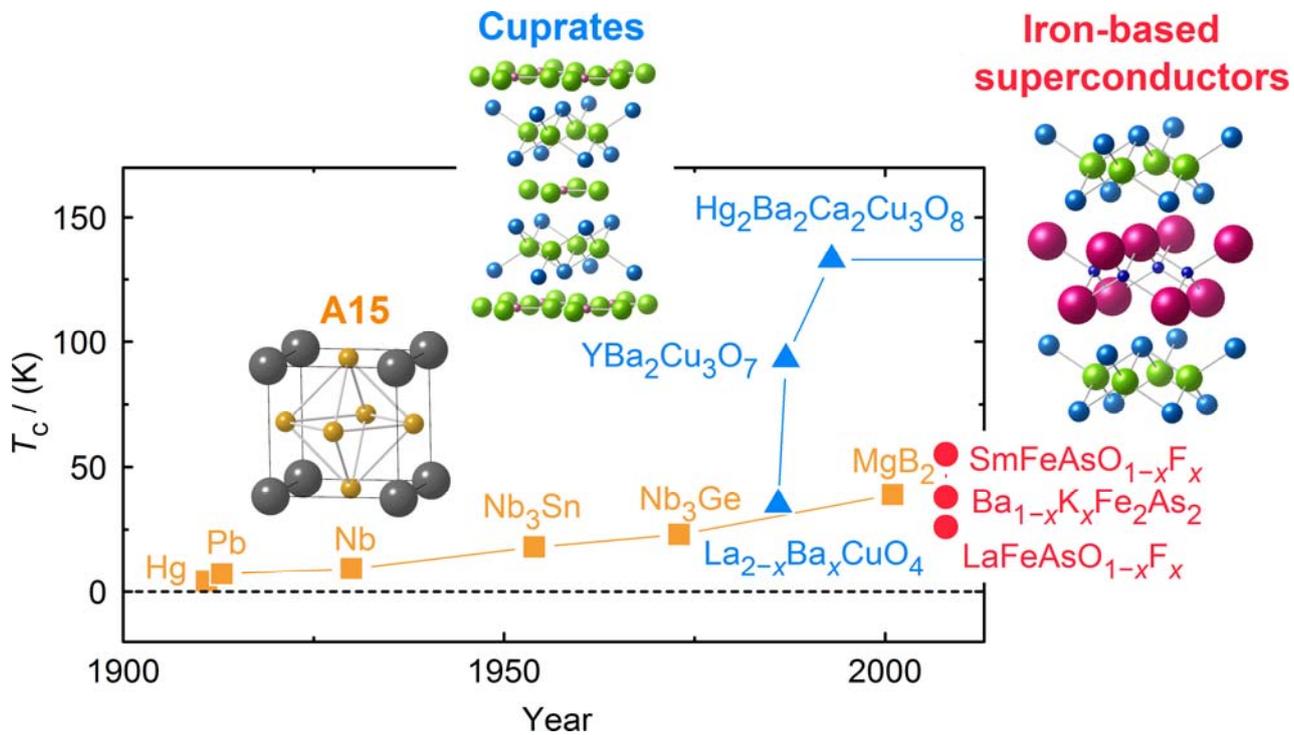

Figure 1. Progress of superconductors. The crystal structures of typical superconductors are shown in the inset.

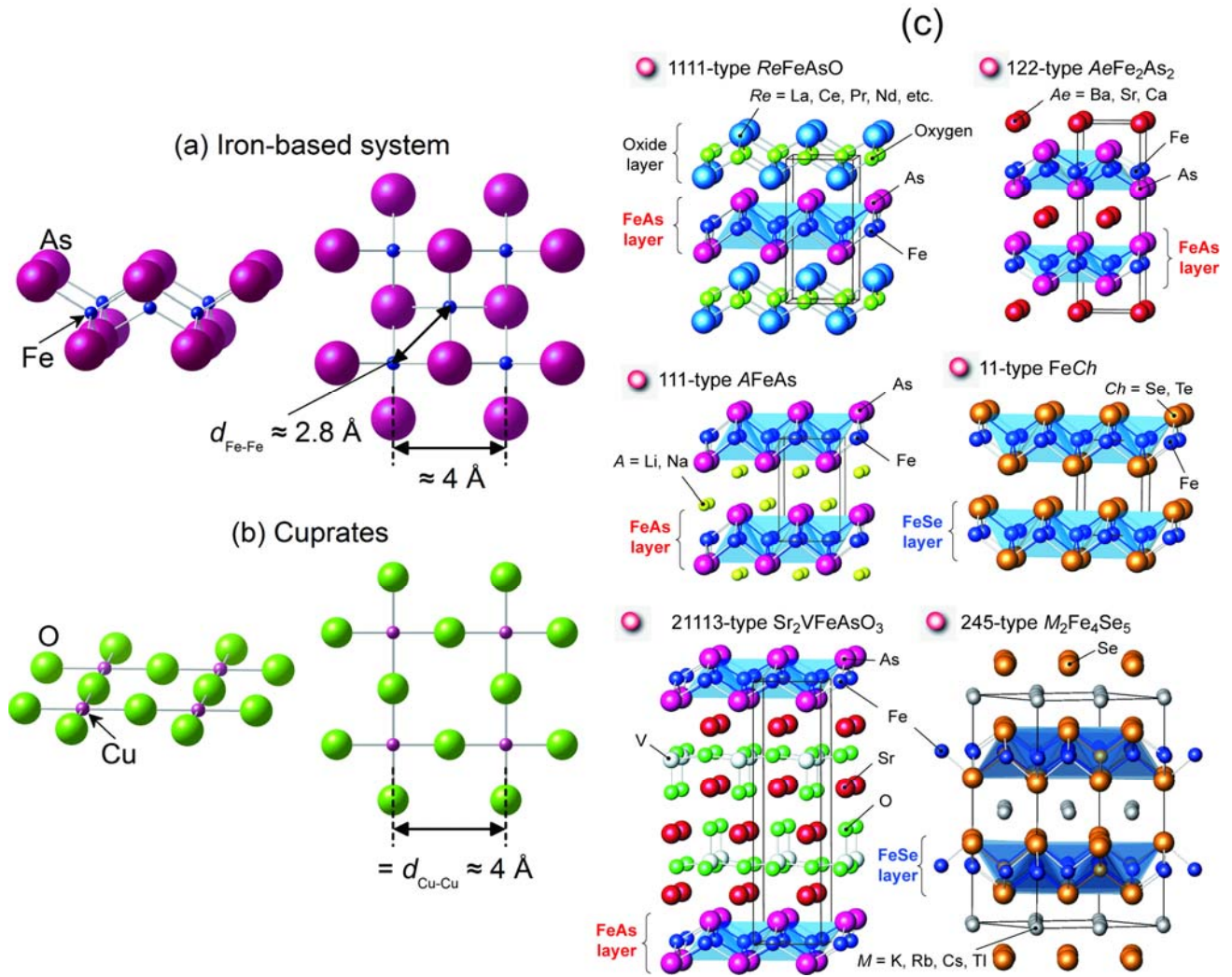

Figure 2. Common structure unit (a) and various parent materials of IBSCs (c). The common structural unit of high-$T_c$ cuprates superconductors (b) is also shown for comparison. These parent compounds are named using an abbreviation of the ratio of the constituent atoms such as "1111" for $Re$FeAsO ($Re$ = rare-earth metal).

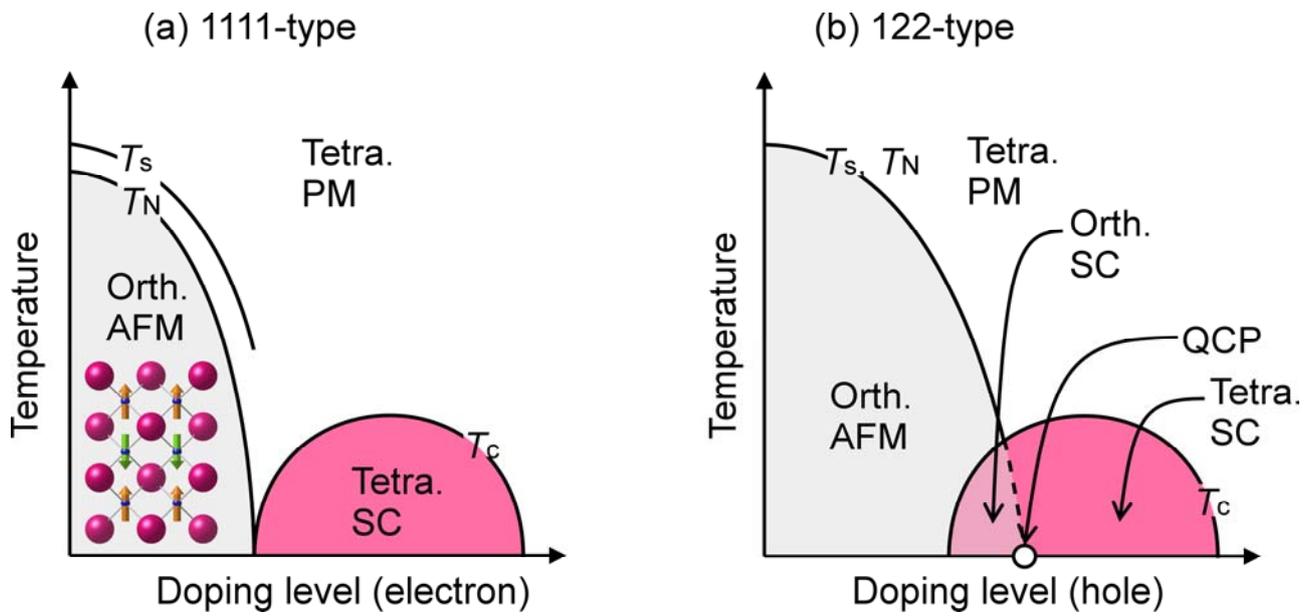

Figure 3. Schematic phase diagram of (a) 1111-type and (b) 122-type systems. $T_s$: structural transition temperature, $T_N$: Néel temperature, SC; superconducting phase, PM: Pauli paramagnetism, QCP: quantum critical point. The distinct differences between the 1111- and 122-type systems are that $T_s$ and $T_N$ are greatly separated in the 1111 system, whereas both are unified in the 122 system, and superconductivity and AFM coexist in the 122 system but not in the 1111 system.

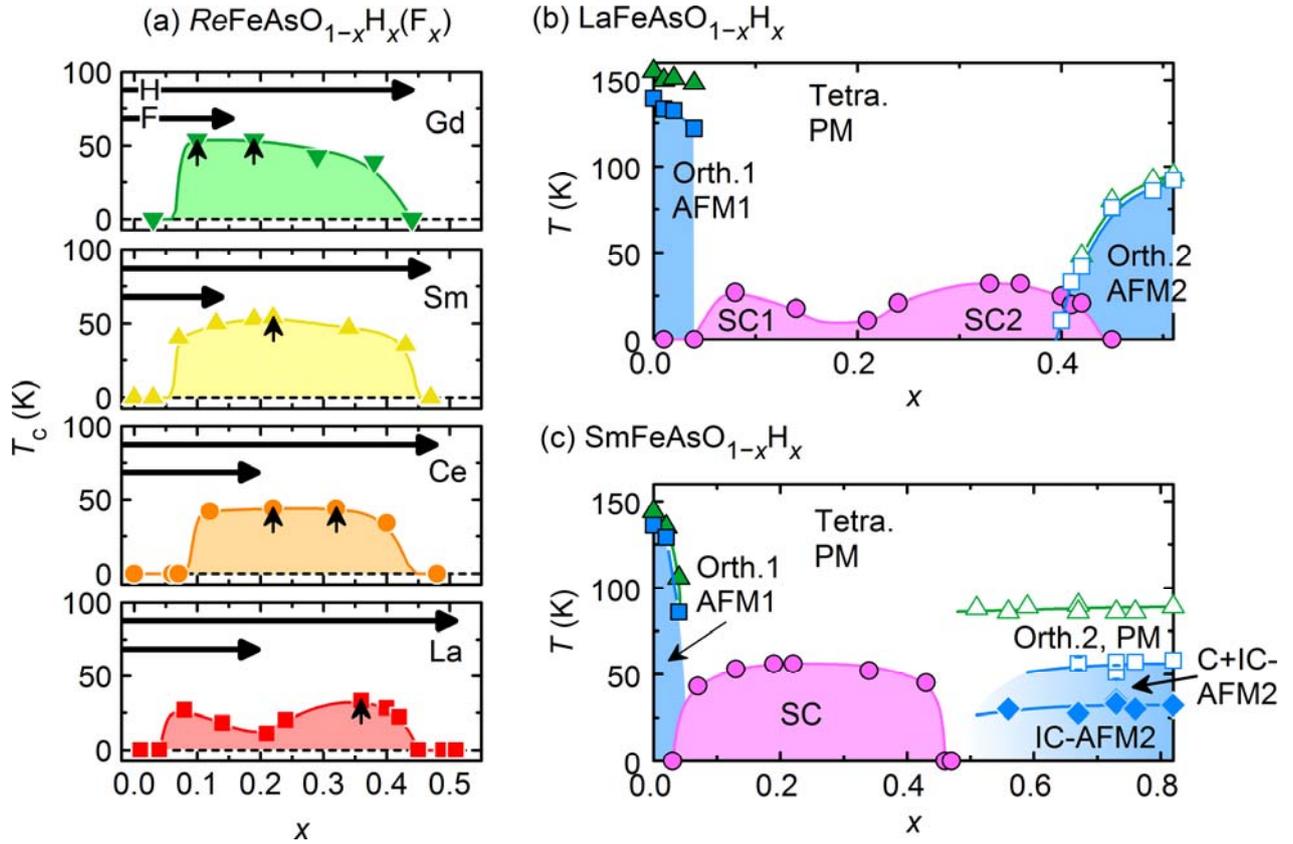

Figure 4. Phase diagram of $Re\text{FeAsO}_{1-x}\text{H}_x$ and comparison between F- and H-substituted $Re$-1111 system ($Re$ = La, Ce, Sm, and Gd) in superconducting region. Two AFM phases are located at the edges of the $T_c$-domes for the La (with double dome) and Sm (with a single dome) systems. Here, AFM: antiferromagnetism, C: commensurate, IC: incommensurate magnetic structure, PM: Pauli paramagnetism.

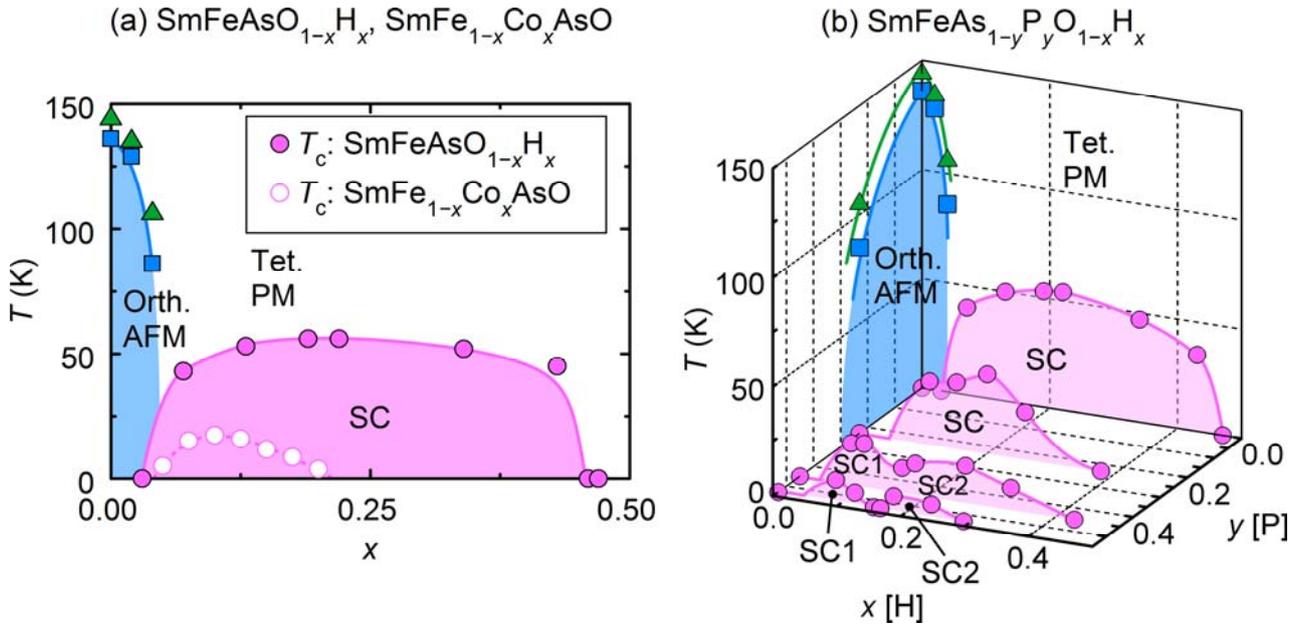

Figure 5. $T_c$ vs. composition in SmFe(Co)AsO$_{1-x}$H$_x$ (a) and SmFeAs$_{1-y}$P$_y$O$_{1-x}$H$_x$ (b).

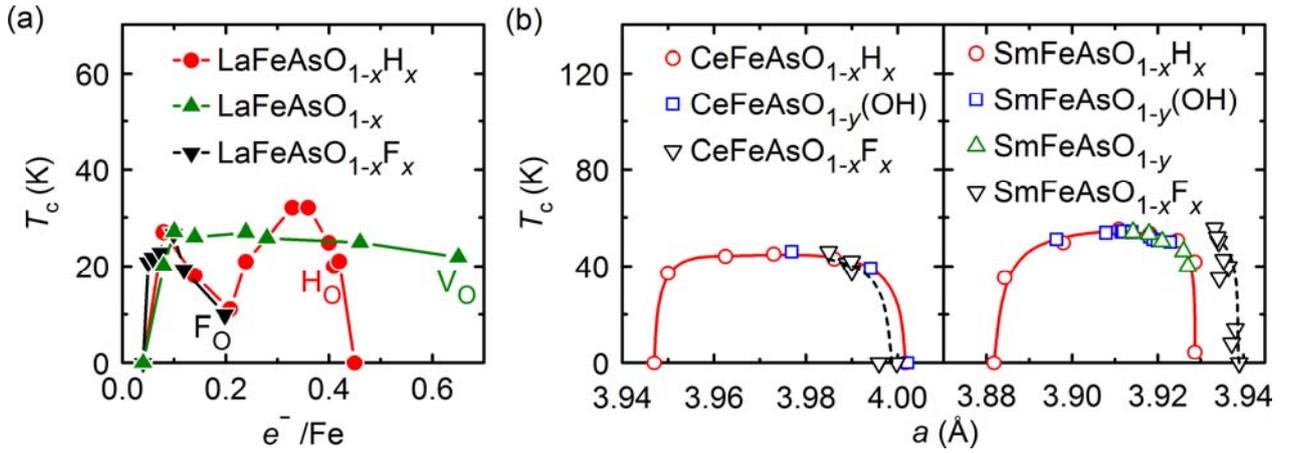

Figure 6. (a) Relationship between observed $T_c$ and doped carrier concentration in LaFeAsO$_{1-x}$H(F)$_x$ and nominal LaFeAsO$_{1-x}$. Here, the doped electron concentration was assumed to be 1 per H(F) or 2 per oxygen vacancy. The observed $T_c$ was obtained from ref. [29] for H doping, ref. [257] for F doping, and ref. [258] for Vo doping. The data on the H-substituted samples (Ho) agree well with the F-substituted ones (Fo); however, the values differed greatly from those on the oxygen-vacancy-doped ones (Vo). (b) Lattice constant vs. $T_c$ in three series of electron-doped Ce/Sm1111 samples using different approaches. These data on nominally oxygen-vacancy-doped and OH-doped samples agree well with those on H-substituted ones. Because only the F-doped samples were synthesized in an ambient atmosphere, the lattice constant was shifted from that of the samples synthesized using high pressure [28].

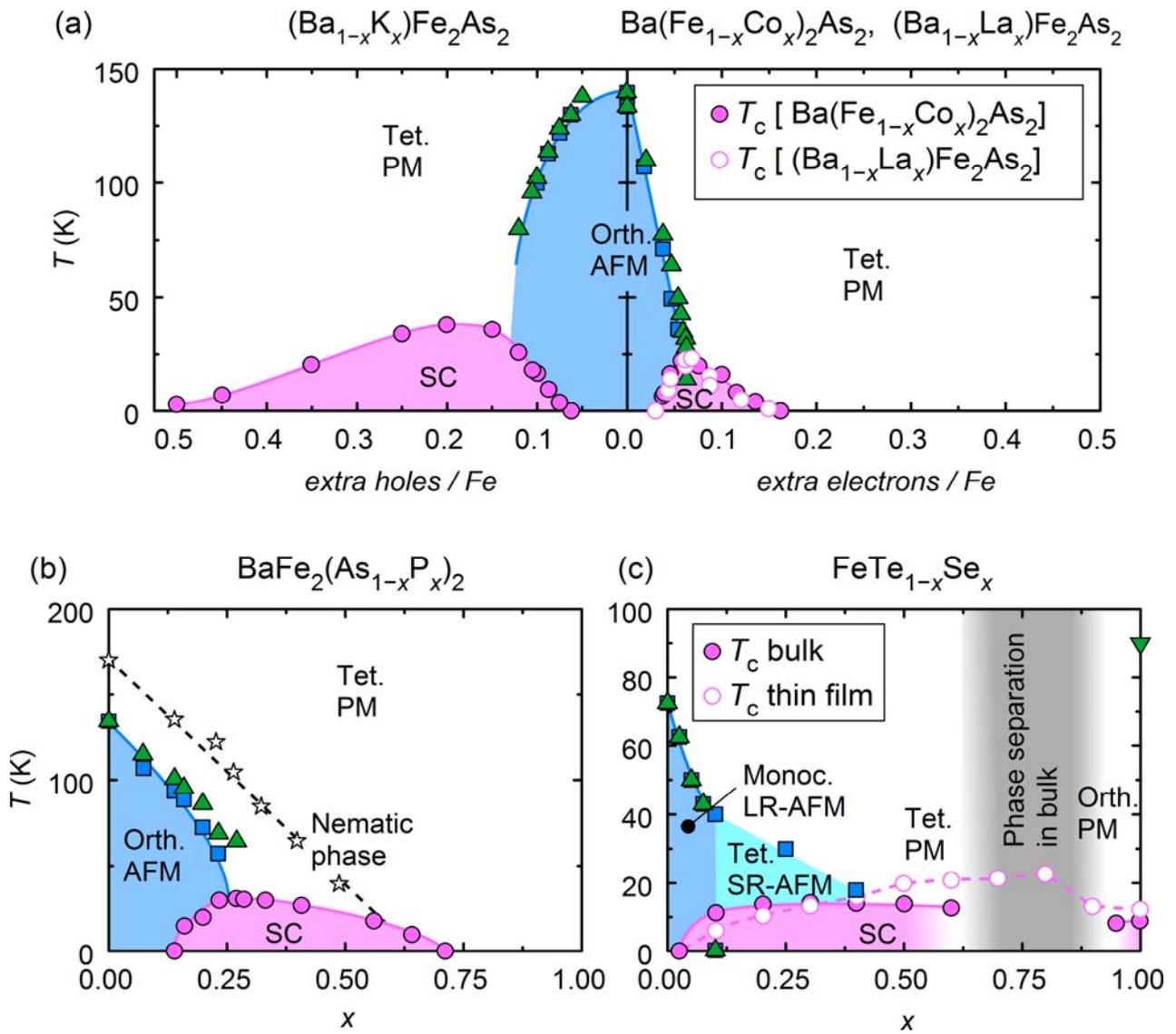

Figure 7. Effect of doping on superconductivity: (a) Hole doping vs. electron doping and (b) and (c) isovalent doping effect.

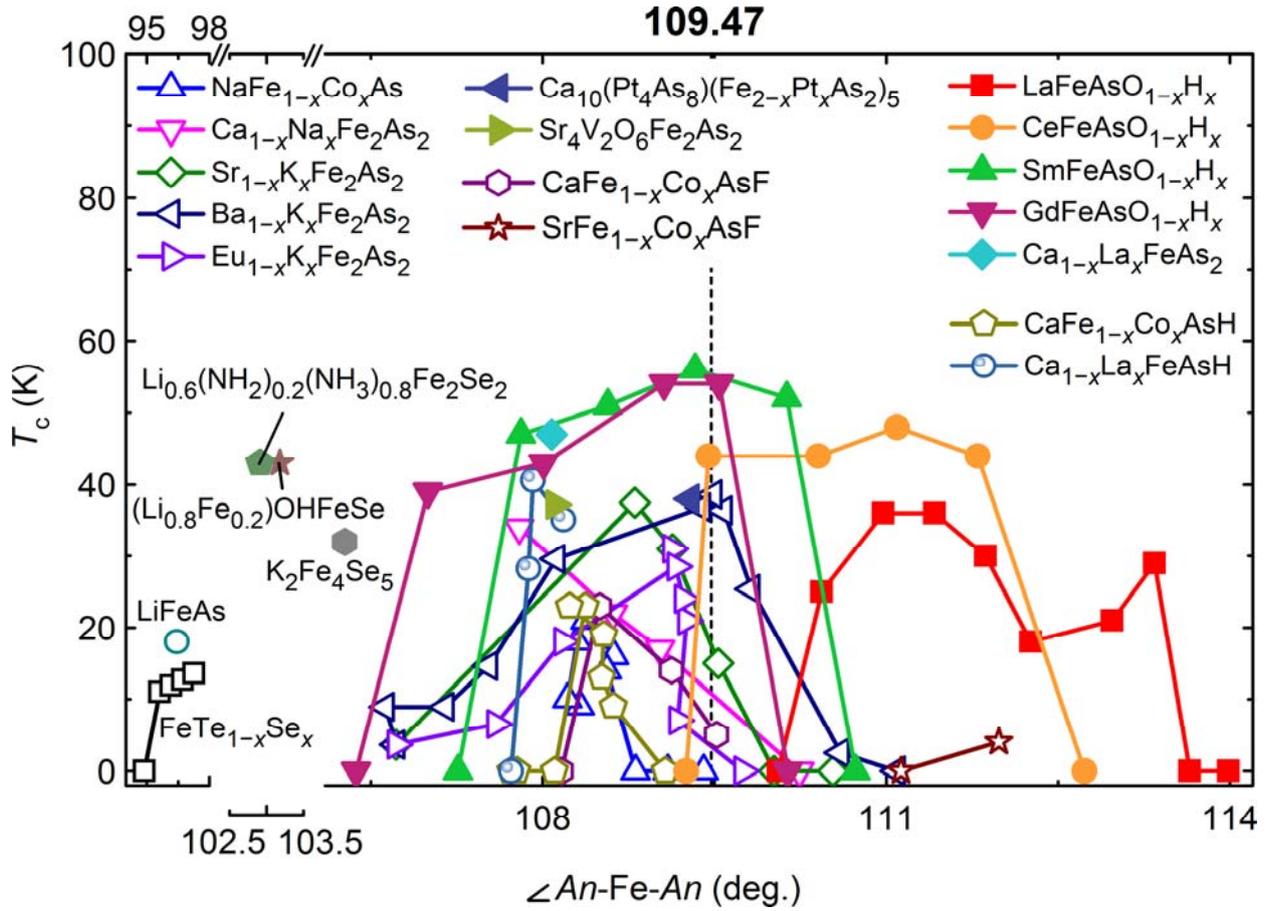

Figure 8. Correlation between $T_c$ and the bond angle of anion–Fe–anion in various IBSCs. The dotted line denotes the bond angle for a regular tetrahedron.

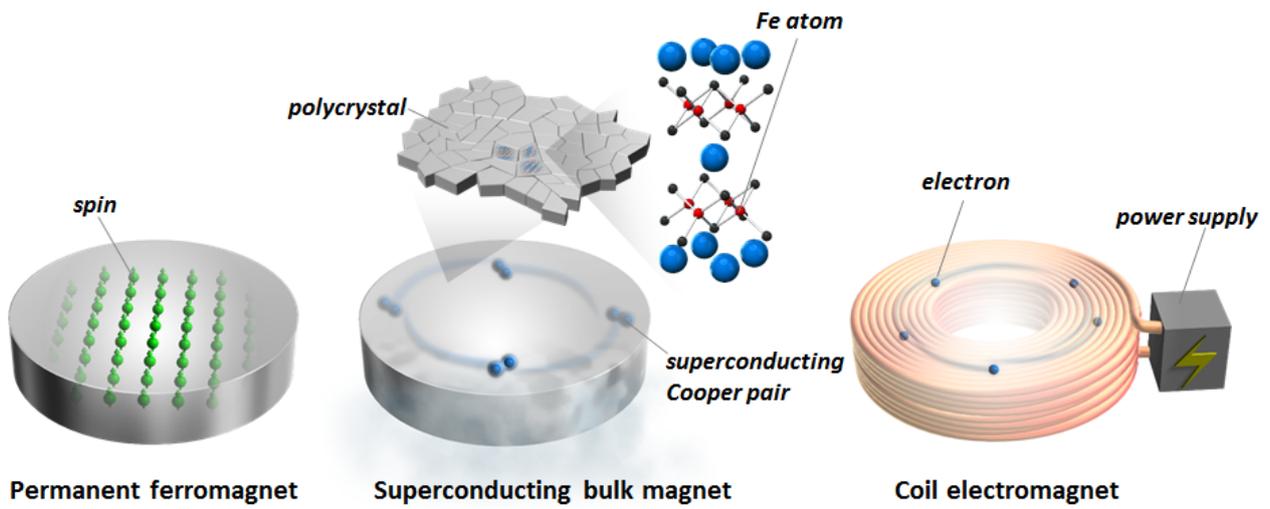

Figure 9. Schematic illustrations of strong magnets: permanent ferromagnet, superconducting bulk magnet, and coil electromagnet.

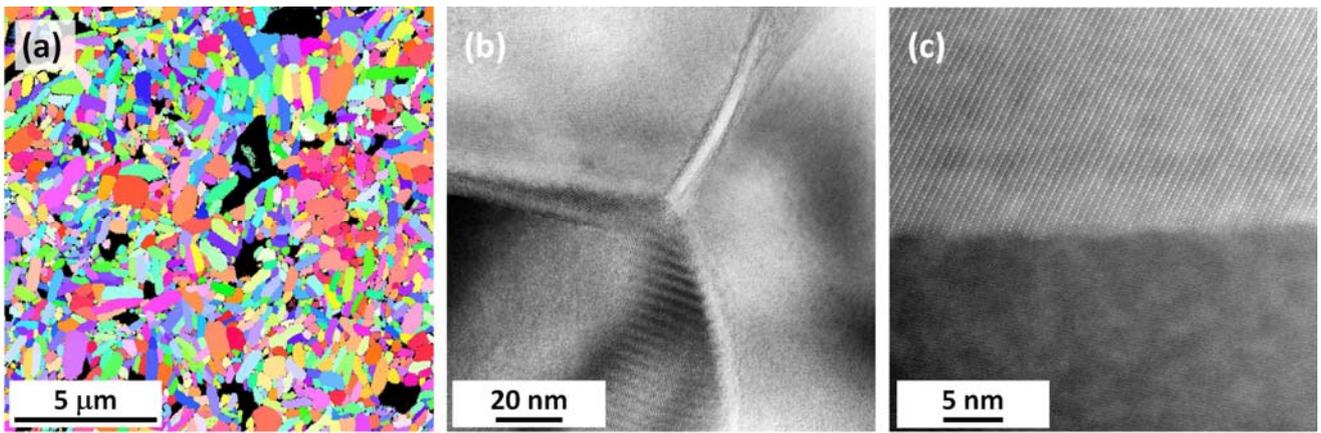

Figure 10. EBSD-IPF image (a) and TEM images (b) and (c) for a Co-doped Ba122 polycrystalline bulk [259].

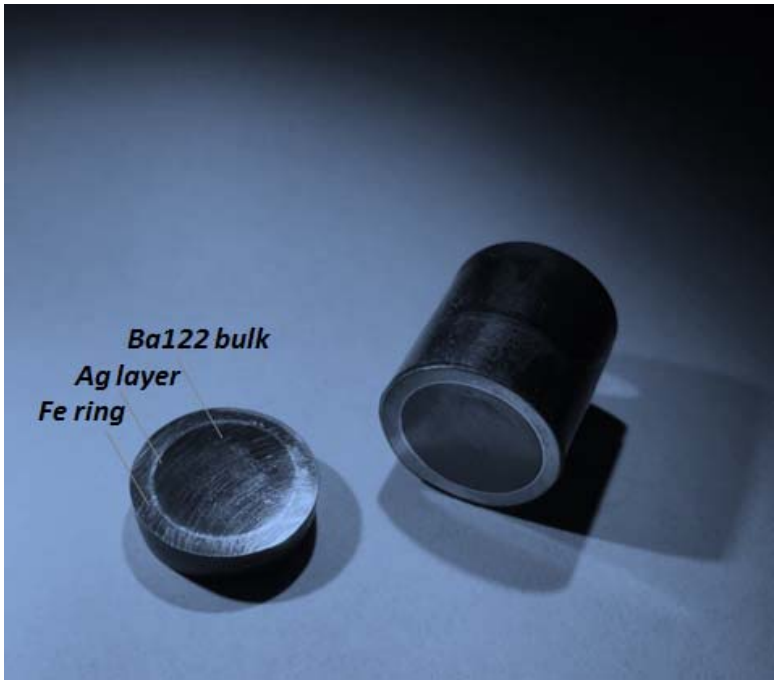

Figure 11. Appearance of polycrystalline K-doped Ba122 bulk [112].

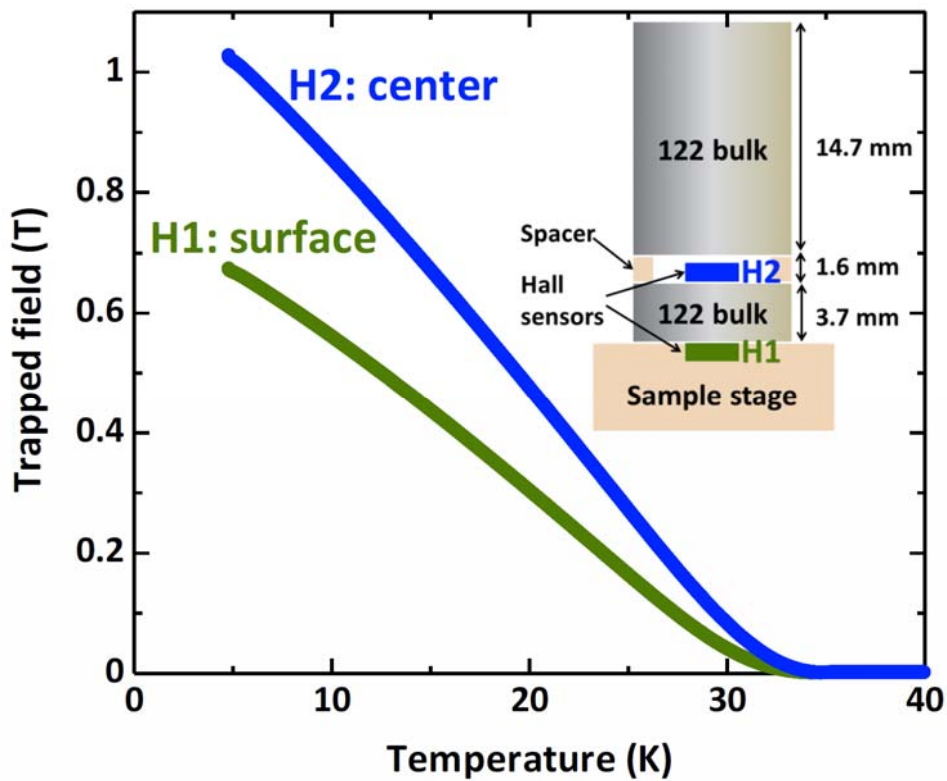

Figure 12. Trapped field as a function of increasing temperature for bulk sample stack field-cooled to ~5 K in 8 T. After reducing the external field to zero, the sample stack was heated, and the relaxation of the trapped field was measured. A schematic of the sample and Hall probe arrangement is shown.[112] Copyright 2015 IOP publishing.

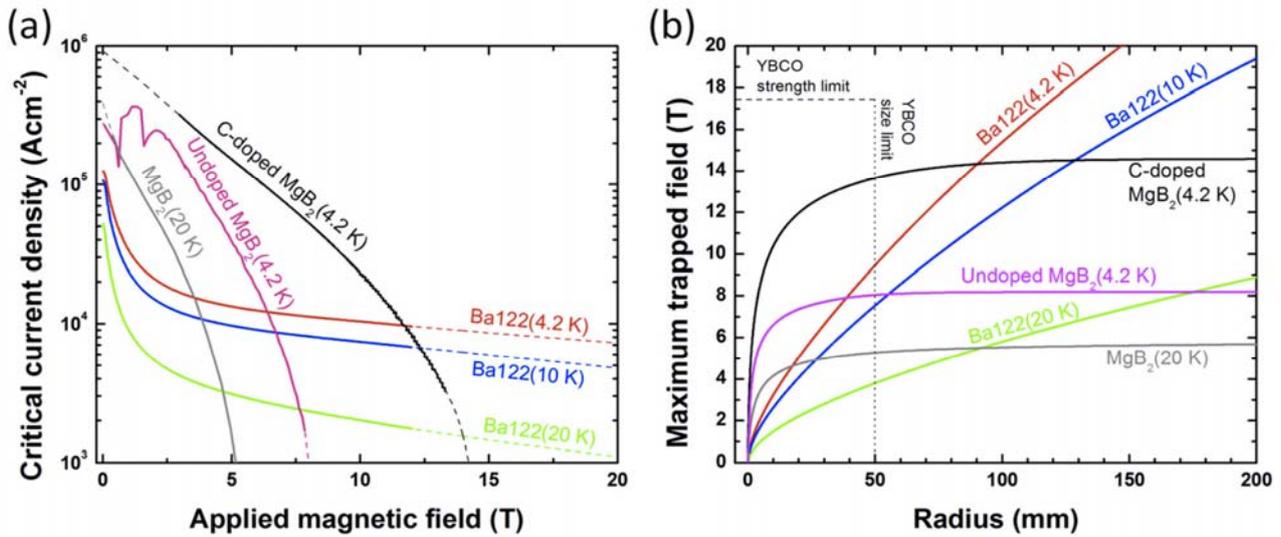

Figure 13. Comparison of K-doped Ba122 and $MgB_2$. (a) Critical current density vs. applied magnetic field for Ba122,[101] undoped $MgB_2$,[260] and C-doped $MgB_2$ bulks.[261] The dotted lines represent extrapolated data. (b) Maximum trapped field vs. radius for K-doped Ba122 and $MgB_2$ polycrystalline bulks calculated from the data in (a) for an infinite thickness cylinder.[112] Copyright 2015 IOP publishing.

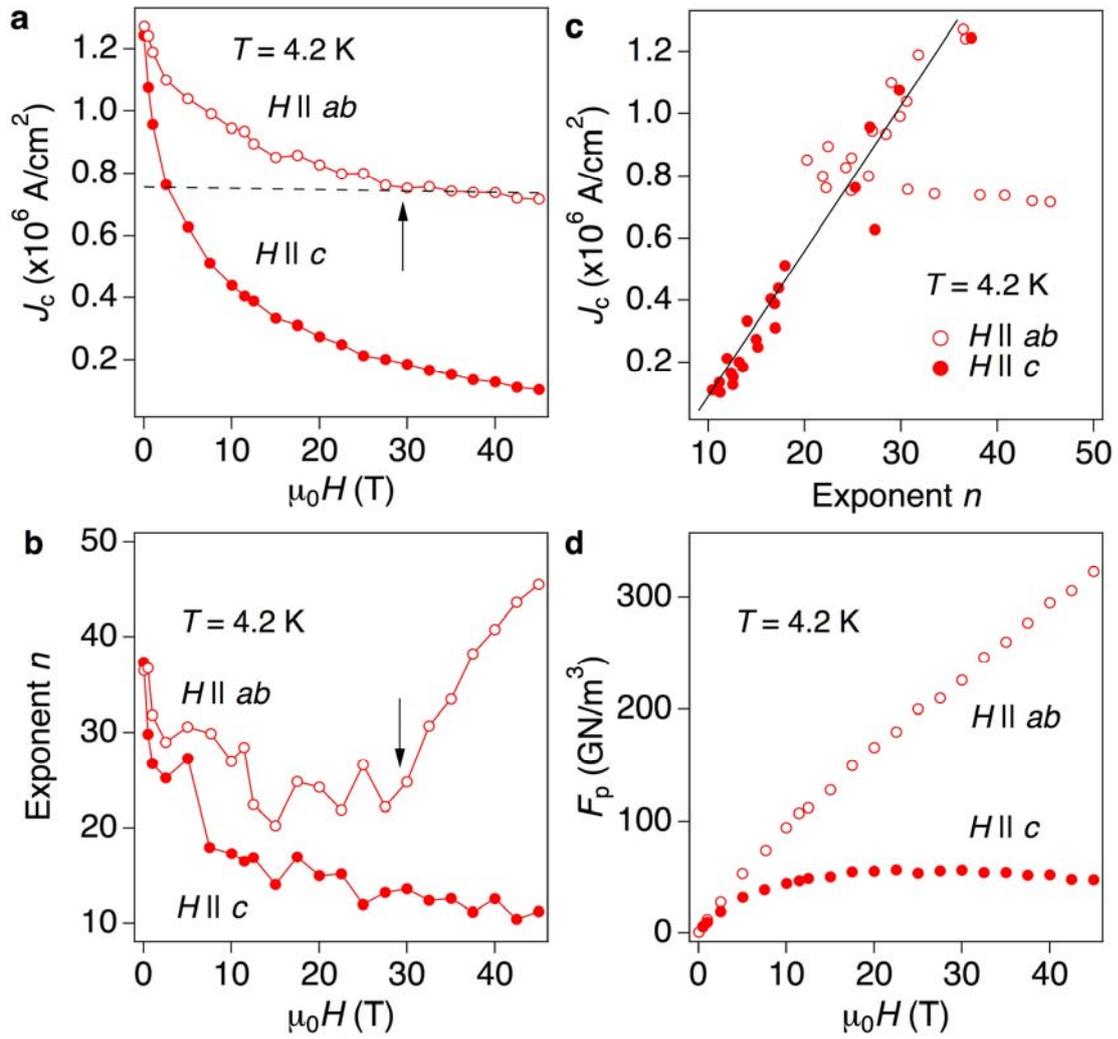

Figure 14. In-field $J_c$ performance of F-doped Sm1111 thin film grown by MBE. (a) Magnetic-field dependence of $J_c$ measured at 4.2 K up to 45 T and (b) corresponding values of the exponent $n$. The crossover from extrinsic to intrinsic pinning is indicated by the arrow. (c) Scaling behavior of the field-dependent $J_c$. (d) Pinning force density $F_p$ at 4.2 K. [136]

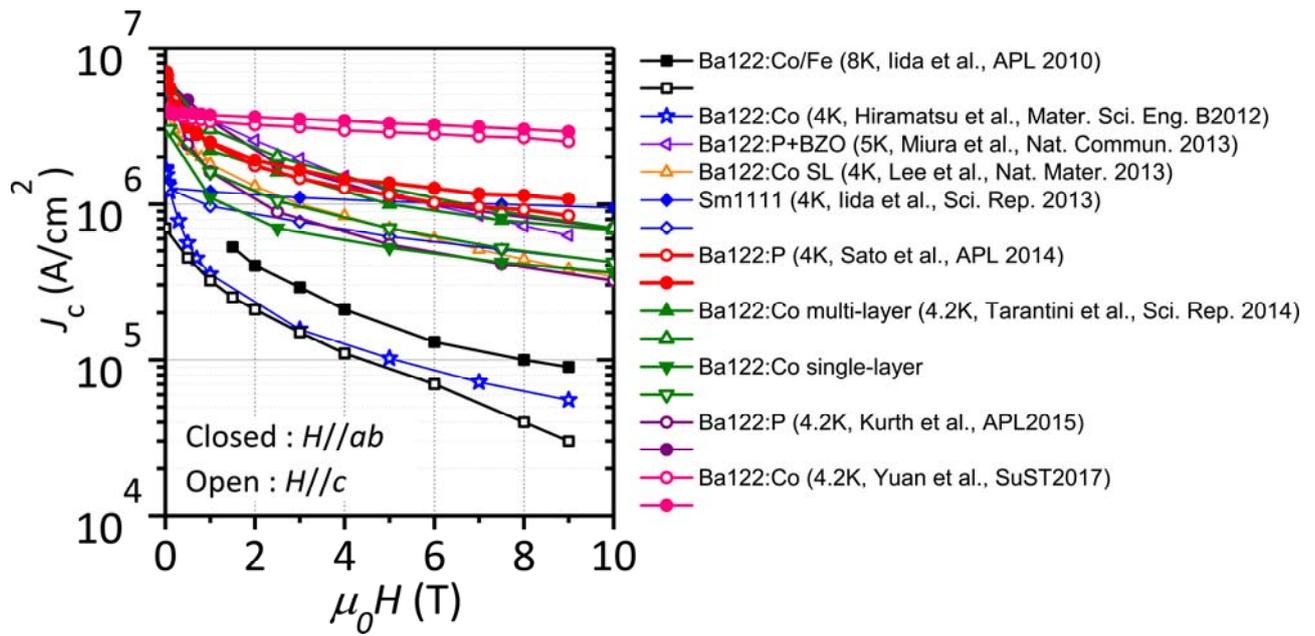

Figure 15. In-field $J_c$ performance for 1111 and 122 epitaxial films at low temperature [262, 263, 159, 136, 164, 156, 165, 147].

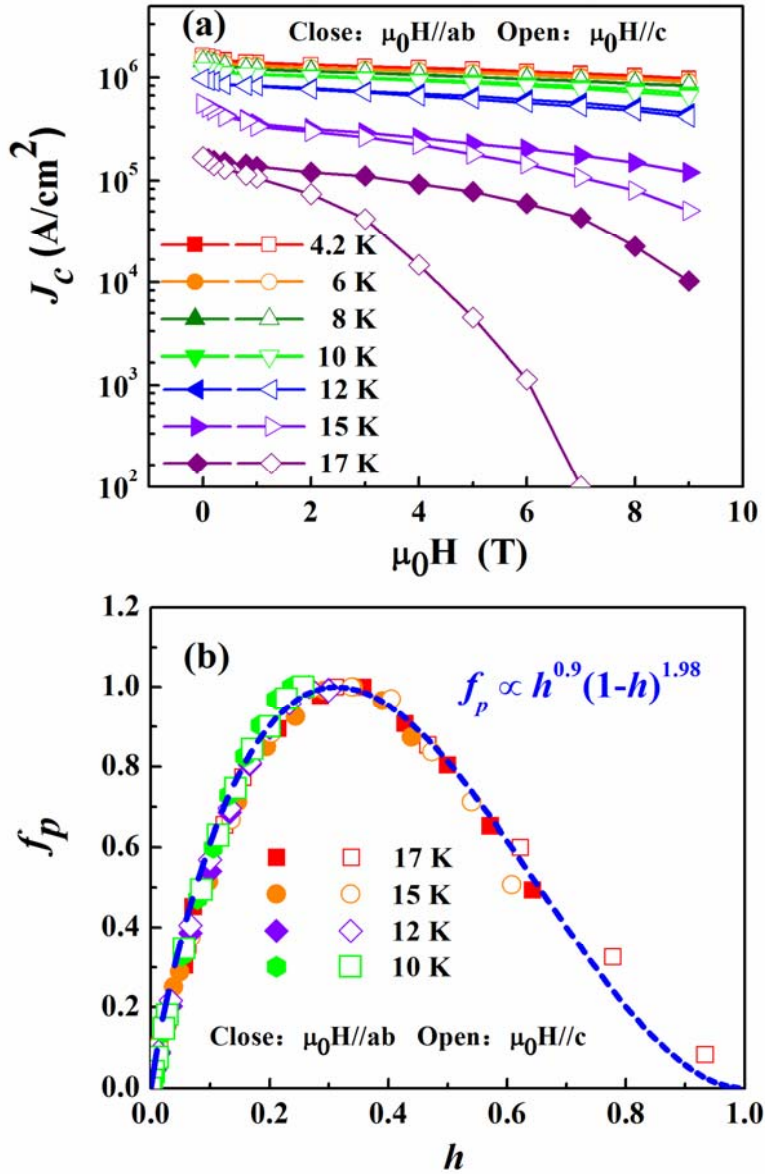

Figure 16. (a) $J_c$ ($H$) of 11 films on a CaF$_2$ substrate at 4.2–17 K for $H//ab$ and $H//c$. (b) Kramer's scaling of pinning force density $f_p$ versus reduced field $h$ at 10–17 K. [174] Copyright 2016 IOP publishing.

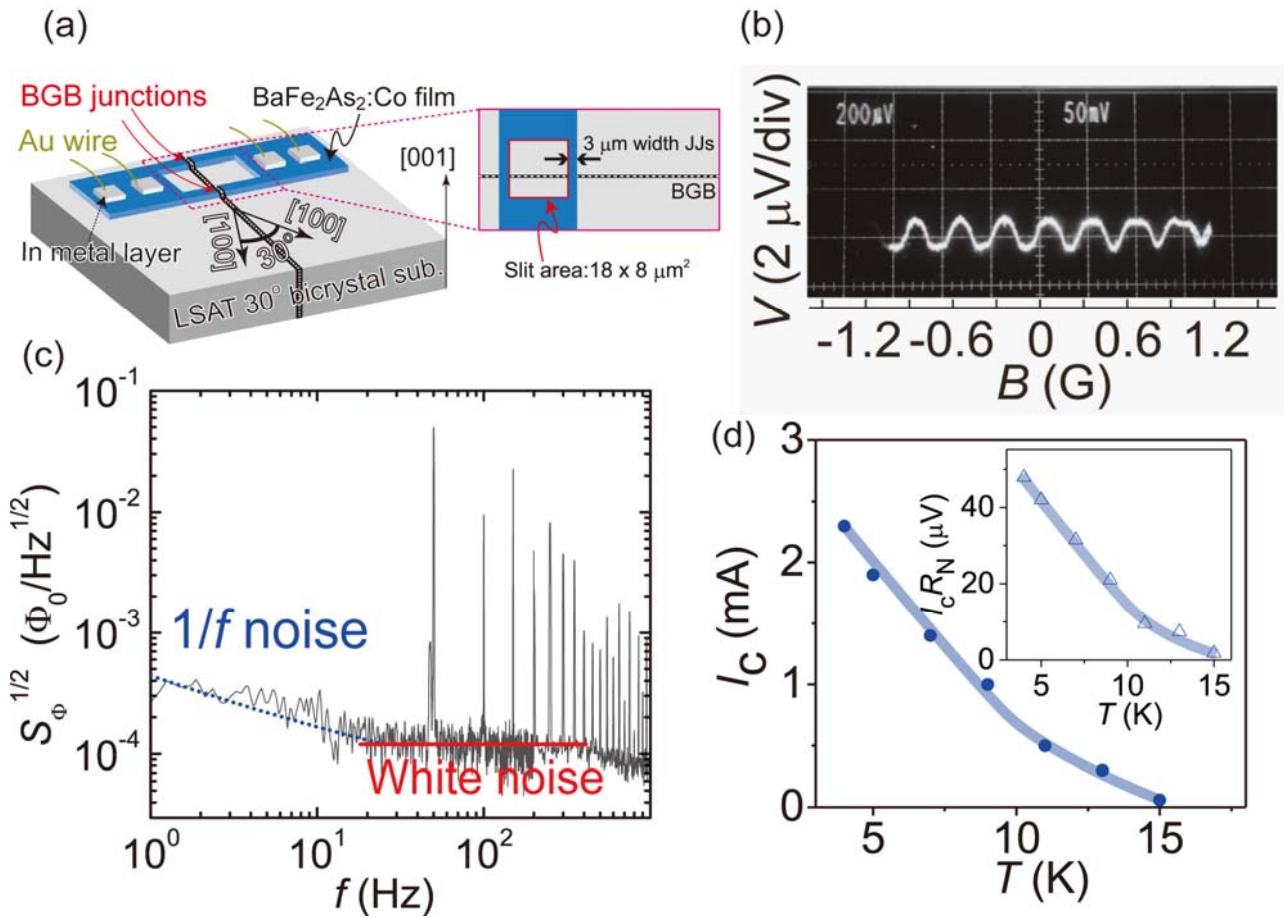

Figure 17. dc-SQUID using Ba122:Co film. (a) Schematic illustration of dc-SQUID consisting of two 3-μm-wide JJs fabricated on [001]-tilt LSAT bicrystal substrate with $\theta_{GB}$ = 30° A Ba122:Co superconducting loop with a slit area of 18 × 8 μm$^2$ was located across the BGB. (b) $V$–$\Phi$ characteristics at 14 K. (c) $S_\Phi^{1/2}$ as a function of $f$ at 14 K. (d) Temperature dependence of $I_c$. The inset shows the $I_cR_N$ product as a function of temperature. [180] Copyright 2010 IOP publishing.

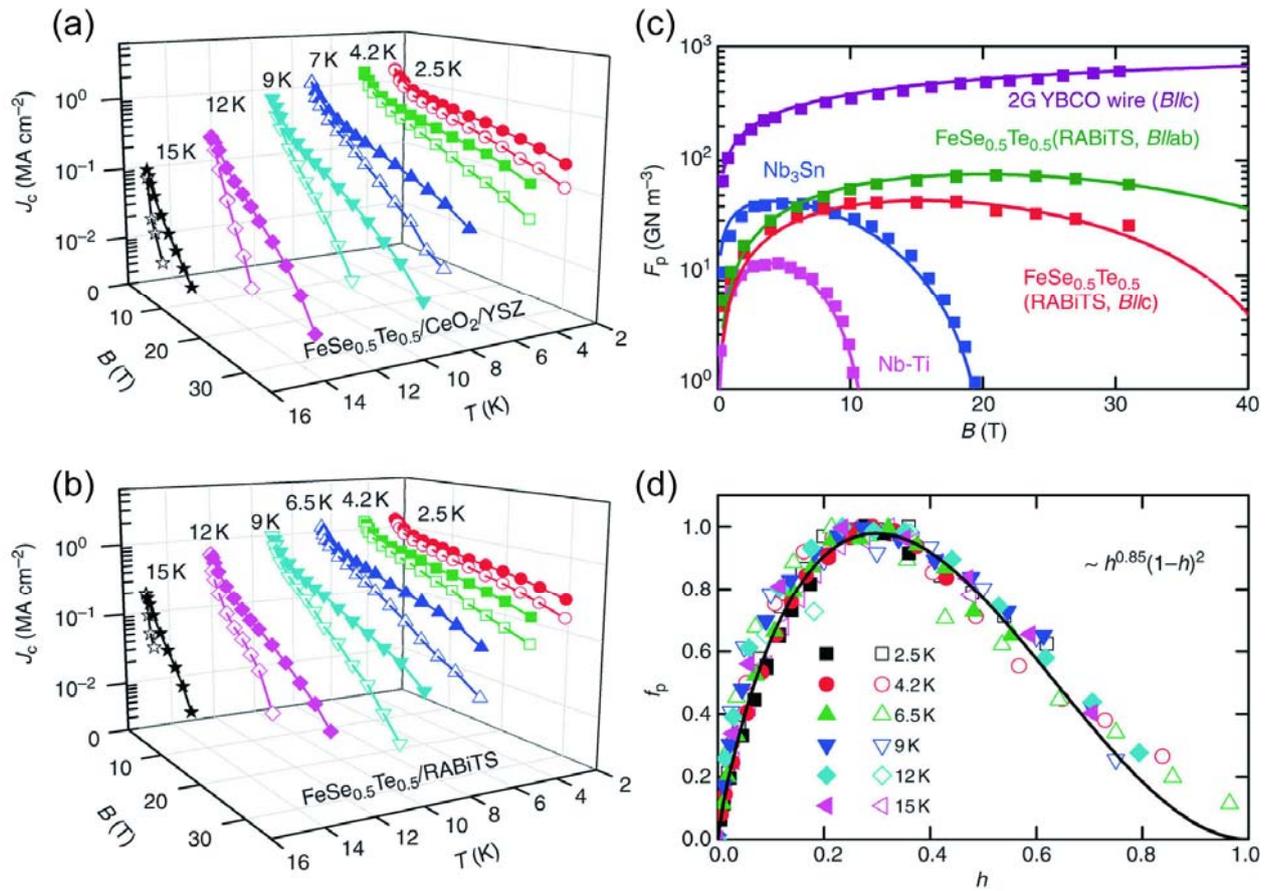

Figure 18. $J_c$ values of 11 films on (a) a YSZ substrate with a $CeO_2$ buffer layer and (b) a RABiTS substrate at various temperatures with the magnetic field parallel (solid symbols) and perpendicular (open symbols) to the ab plane (tape surface). The self-field $J_c$ of both films were above 1 MA/cm$^2$ at 4.2 K. Under 30-T magnetic fields, both films still exhibited $J_c$ of ~0.1 MA/cm$^2$. (c,d) Pinning force analysis for 11 film grown on RABiTS. (a) $F_p$ at 4.2 K of 11 film grown on RABiTS substrate. (b) Scaling of pinning force density versus reduced field $h$ for 11 film grown on RABiTS substrate at various temperatures with field perpendicular (solid symbols) and parallel (open symbols) to the *c*-axis. [186] Copyright 2013 Nature Publishing Group

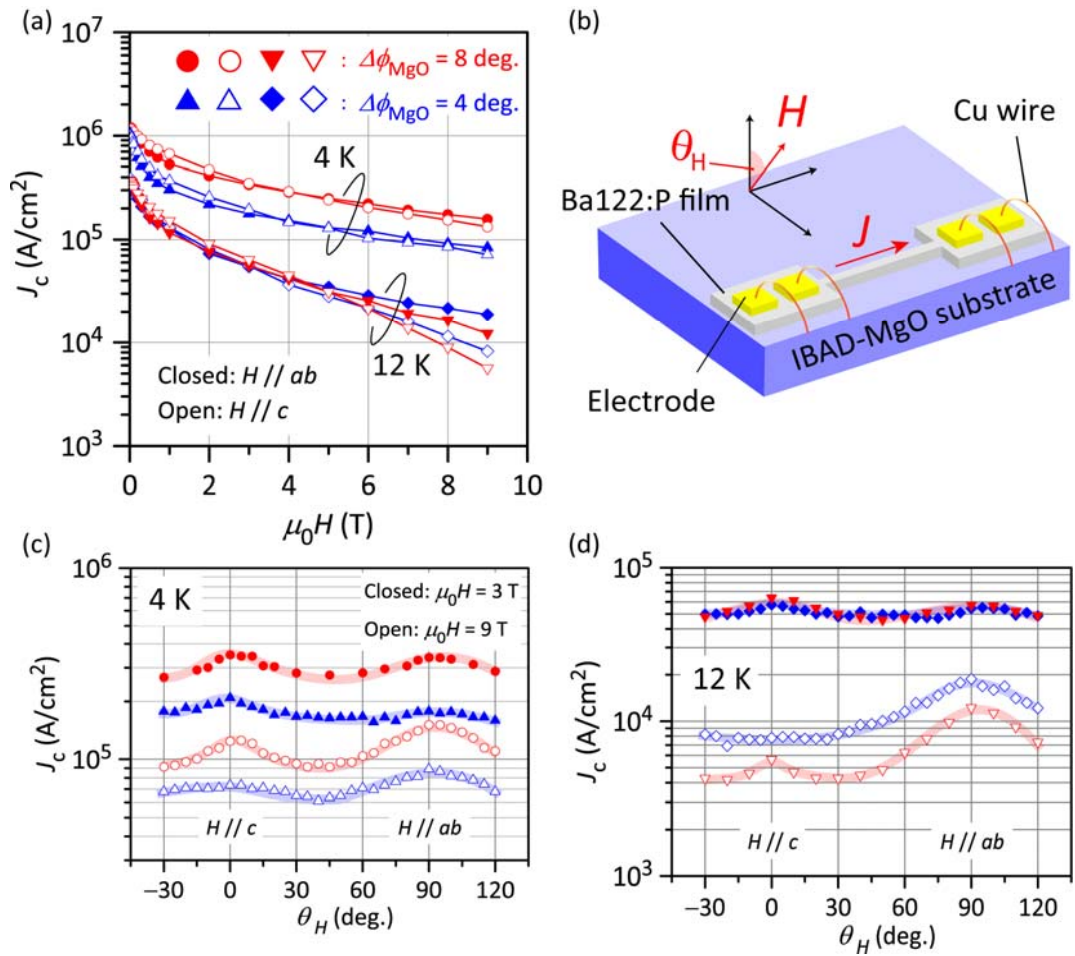

Figure 19. $J_c$ values of Ba122:P coated conductors on two types of IBAD metal-tape substrates with $\Delta\phi_{MgO}$ = 8° (poorly aligned, red symbols) and 4° (well aligned, blue symbols). (a) $H$ dependence of $J_c$ at 4 K and 12 K. The closed and open symbols indicate the configurations of the $H \parallel ab$ plane and $H \parallel c$ axis of the Ba122:P films, respectively. (b) Relationship between $J$ and $H$ directions under $J_c$ measurement. $\theta_H$ was varied from −30° to 120°. (c,d) $\theta_H$ dependence of $J_c$ at (c) 4 K and (d) 12 K. The closed and open symbols represent the data for $\mu_0 H$ = 3 and 9 T, respectively. [188]

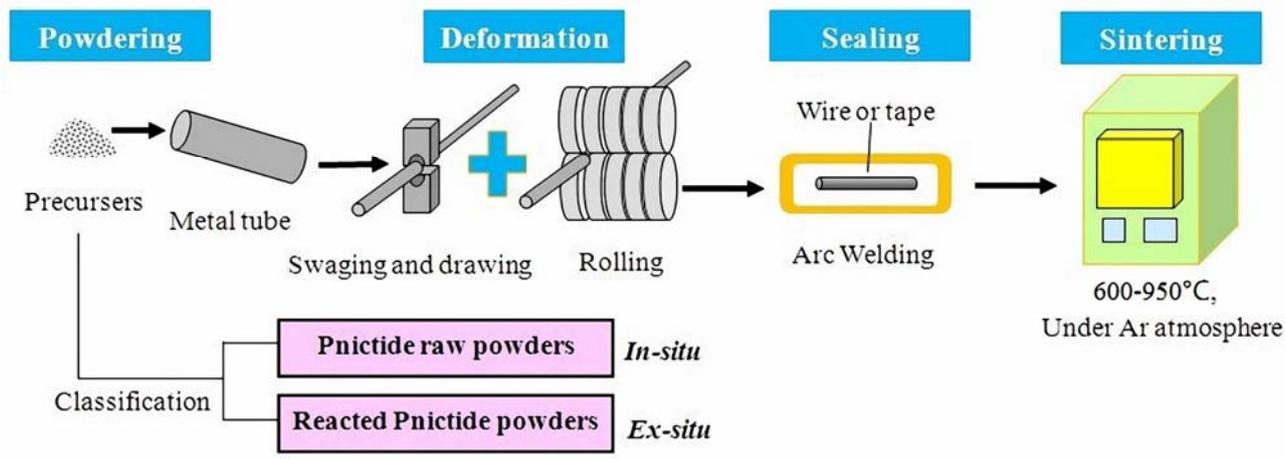

Figure 20. PIT process used for fabricating IBSC wires and tapes.

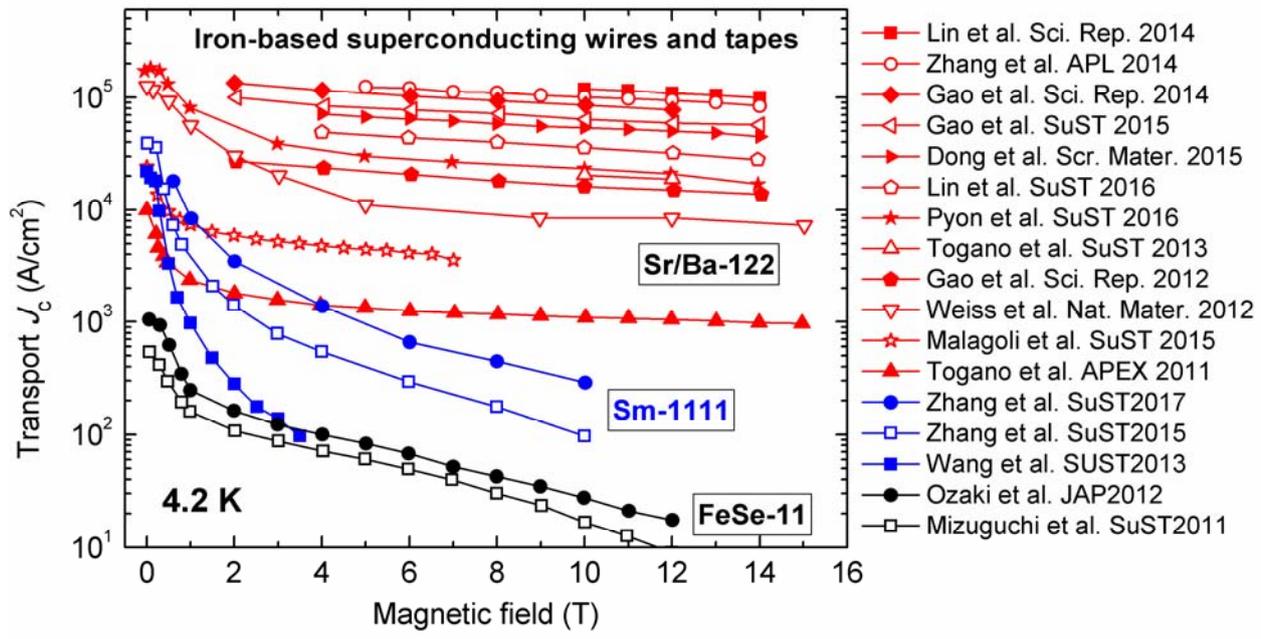

Figure 21. Transport $J_c$ at 4.2 K as a function of applied magnetic field for PIT-processed 122, 1111, and 11 wires and tapes. The data refer to Sr/Ba-122 wires and tapes [211, 242, 241, 209, 225, 253, 235, 240, 206, 100, 224, 202], 1111 tapes [228, 97, 227], and 11 wires [213, 212].

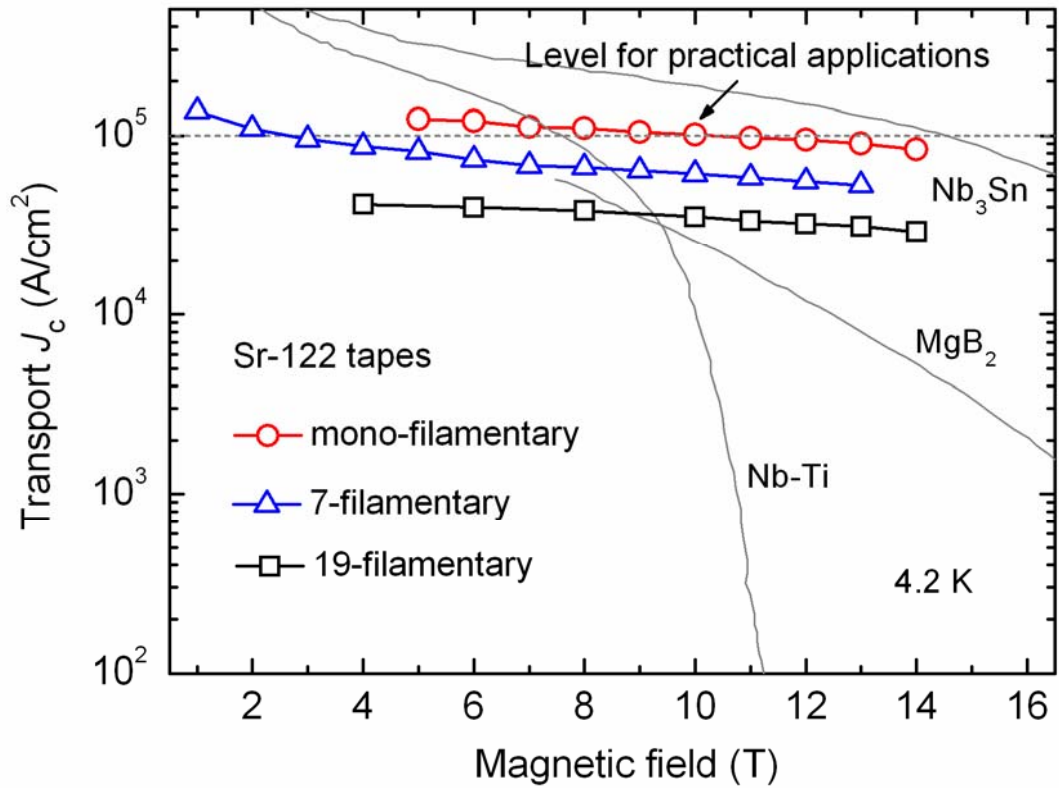

Figure 22. $J_c$ as a function of applied field for hot-pressed mono- and multi-filamentary Sr-122 tapes. Values for $MgB_2$ and commercial NbTi and $Nb_3Sn$ wires have been added for comparison.

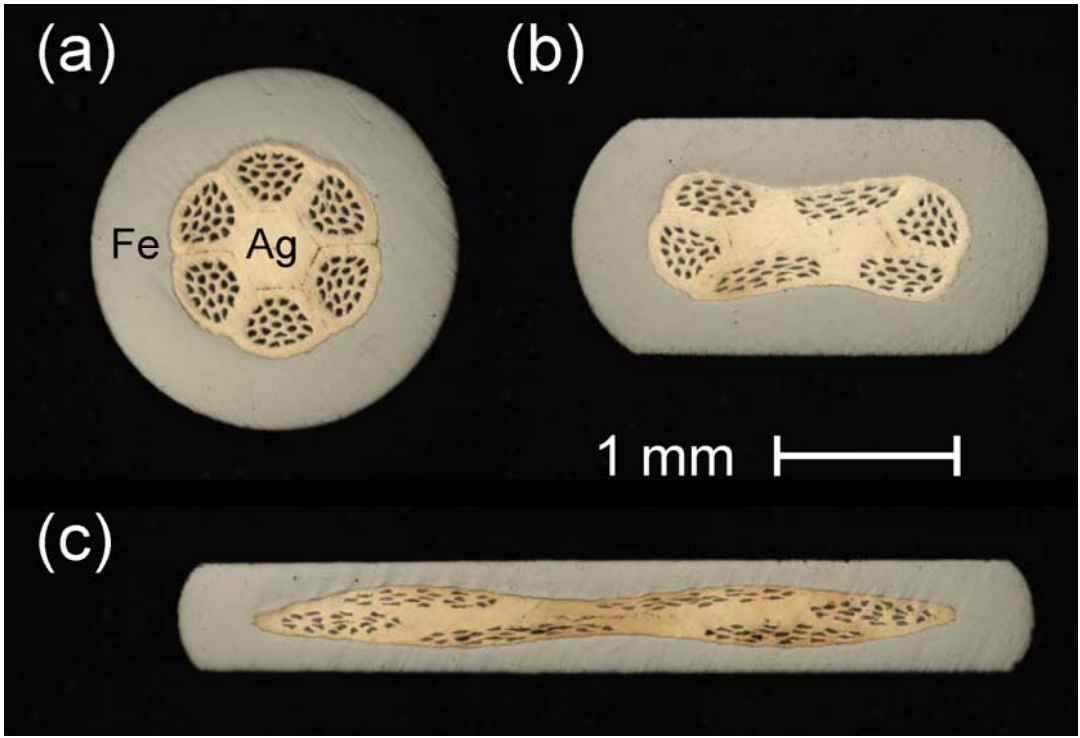

Figure 23. Cross-section of 114 multi-filamentary 122-type IBSC wires and tapes fabricated using a scalable PIT process[248]. Copyright 2015 AIP publishing

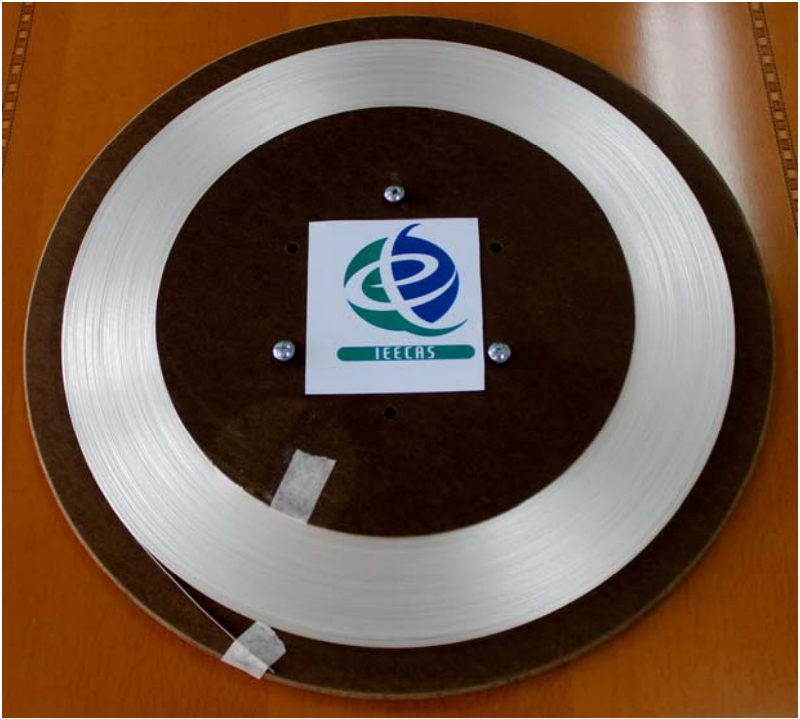
Figure 24. The first 100-m-class 122-type IBSC wire developed by the Institute of Electrical Engineering, Chinese Academy of Sciences (IEECAS).

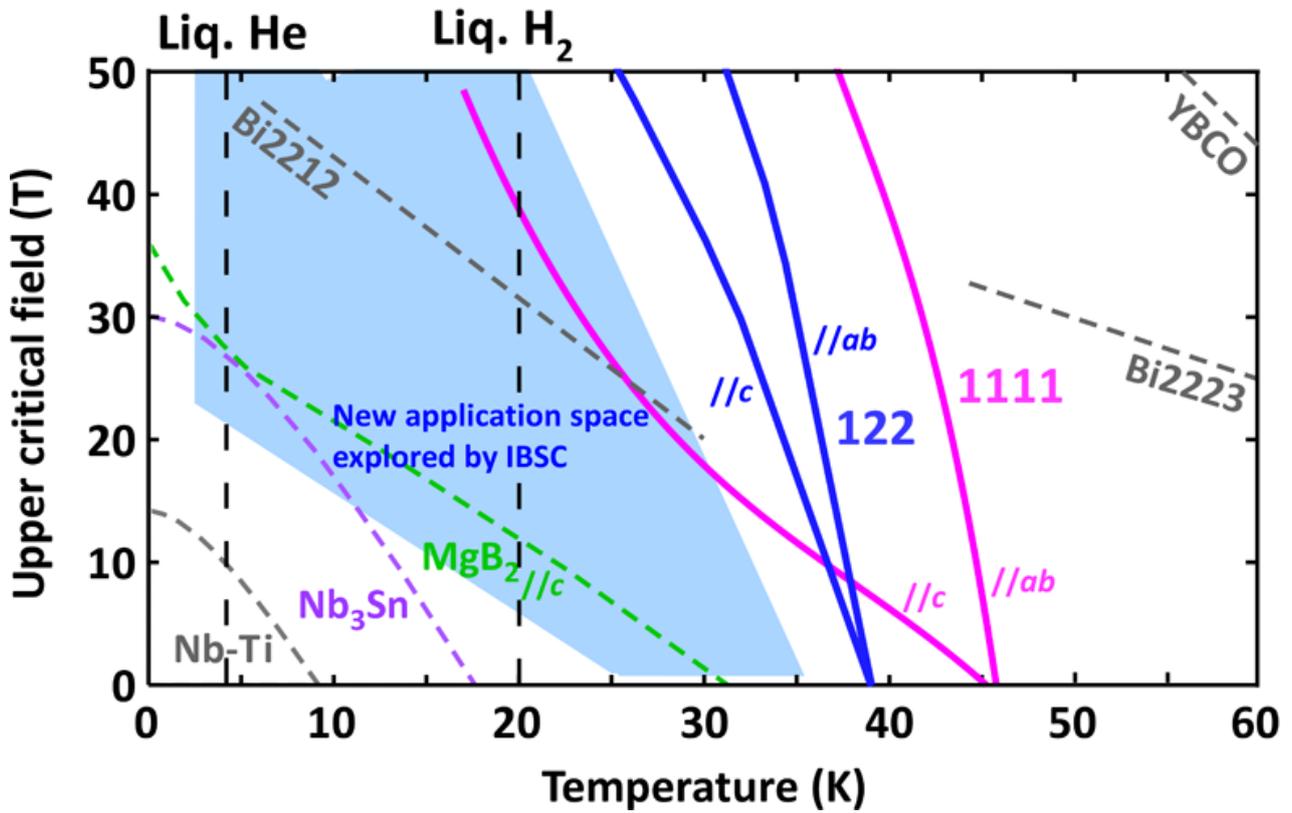

Figure 25. Comparison of upper critical field $H_{c2}$ for various high-field superconductors, including metallic Nb–Ti, Nb$_3$Sn, and MgB$_2$ [264]; the IBSCs NdFeAsO$_{1-x}$F$_x$ (1111) [64] and Ba$_{1-x}$K$_x$Fe$_2$As$_2$ (122) [115]; and cuprate superconductors YBCO [265], Bi2212 [265], and Bi2223 [266]. For the cuprates, 50% of the in-field resistive transition was regarded as $H_{c2}^{//c}$. The light blue area highlights the temperature and field range in which the application of IBSCs would be most effective. The vertical broken lines indicate the boiling points at 1 atm of the cryogenic coolants liquid helium and liquid hydrogen.